\documentclass[fleqn,usenatbib]{mnras}

\usepackage{newtxtext,newtxmath}

\usepackage[T1]{fontenc}
\usepackage{lineno}

\DeclareRobustCommand{\VAN}[3]{#2}
\let\VANthebibliography\thebibliography
\def\thebibliography{\DeclareRobustCommand{\VAN}[3]{##3}\VANthebibliography}

\newcommand{\hinvMsun}{h^{-1}M_{\odot}}
\newcommand{\hinvsqLsun}{h^{-2}L_{\odot}}
\newcommand{\hinvMpc}{h^{-1}\,{\rm Mpc}}

\newcommand{\kms}{{\rm km\,s}^{-1}}
\newcommand{\avg}[1]{\langle #1 \rangle}

\usepackage{graphicx}	
\usepackage{amsmath}	
\usepackage{subfigure}

\graphicspath{{./plots/}}

\title[Subaru HSC weak lensing mass observable scaling relations of GAMA groups]{The Subaru HSC weak lensing mass-observable scaling relations of spectroscopic galaxy groups from the GAMA survey}

\author[Rana et al.]{
Divya Rana,$^{1}$\thanks{E-mail: divyar@iucaa.in}
Surhud More,$^{1,2}$\thanks{E-mail: surhud@iucaa.in}
Hironao~Miyatake,$^{3,2,4}$
Takahiro~Nishimichi,$^{7,2}$
\newauthor
Masahiro~Takada,$^{2}$
Aaron S. G. Robotham,$^{8}$
Andrew M. Hopkins $^{9}$
\newauthor
and Benne W. Holwerda$^{10}$
\\
$^{1}$Inter University Centre for Astronomy and Astrophysics, Ganeshkhind, Pune 411007, IN\\
$^{2}$Kavli Institute for the Physics and Mathematics of the Universe (WPI), University of Tokyo, 5-1-5, Kashiwanoha, 2778583, JP\\
$^{3}$Kobayashi-Maskawa Institute for the Origin of Particles and the Universe (KMI), Nagoya University, Nagoya, 464-8602, JP\\
$^{4}$Institute for Advanced Research, Nagoya University, Nagoya 464-8601, JP\\
$^{5}$Division of Particle and Astrophysical Science, Graduate School of Science, Nagoya University, Nagoya 464-8602, JP\\
$^{6}$Jet Propulsion Laboratory, California Institute of Technology,
Pasadena, CA 91109, USA\\
$^{7}$Center for Gravitational Physics, Yukawa Institute for Theoretical Physics, Kyoto University, Kyoto 606-8502, JP\\
$^{8}$ICRAR, M468, University of Western Australia, Crawley, WA 6009, Australia \\
$^{9}$ Australian Astronomical Optics, Macquarie University, 105 Delhi Rd, North Ryde, NSW 2113, Australia \\
$^{10}$ Department of Physics and Astronomy, 102 Natural Science Building, University of Louisville, Louisville KY 40292, USA
}

\date{Accepted XXX. Received YYY; in original form ZZZ}

\pubyear{2021}

\begin{document}
\label{firstpage}
\pagerange{\pageref{firstpage}--\pageref{lastpage}}
\maketitle

\begin{abstract}
We utilize the galaxy shape catalogue from the first-year data release of the Subaru Hyper Suprime-cam Survey (HSC) to study the dark matter content of galaxy groups in the Universe using weak lensing. We use galaxy groups from the Galaxy Mass and Assembly galaxy survey in approximately $100$ sq. degrees of the sky that overlap with the HSC survey as lenses. We restrict our analysis to the $1587$ groups with at least five members. We divide these groups into six bins each of group luminosity and group member velocity dispersion and measure the lensing signal with a signal-to-noise ratio of $55$ and $51$ for these two different selections, respectively. We use a Bayesian halo model framework to infer the halo mass distribution of our groups binned in the two different observable properties and constrain the power-law scaling relation, and the scatter between mean halo masses and the two group observable properties. We obtain a 5 percent constraint on the amplitude of the scaling relation between halo mass and group luminosity with $\avg{M} = (0.81\pm 0.04)\times10^{14}\hinvMsun$ for $L_{\rm grp}=10^{11.5}\hinvsqLsun$, and a power-law index of $\alpha=1.01\pm 0.07$. We constrain the amplitude of the scaling relation between halo mass and velocity dispersion to be $\avg{M}=(0.93\pm 0.05)\times10^{14}\hinvMsun$ for $\sigma=500 \kms$ and a power-law index to be $\alpha=1.52\pm0.10$. However, these scaling relations are sensitive to the exact cuts applied to the number of group members. Comparisons with similar scaling relations from the literature show that our results are consistent and have significantly reduced errors.
\end{abstract}




\section{Introduction}

Structure formation in the Universe proceeds hierarchically, where the lowest mass haloes form first and subsequently merge with each other to form more massive dark matter haloes \citep[][]{2012ARA&A..50..353K}. There is much theoretical progress in the understanding of the formation of dark matter haloes, especially with the help of numerical simulations \citep[for a review see][]{2012AnP...524..507F}. However, our understanding of the processes which result in the formation of galaxies within these dark matter haloes remains relatively less understood \citep[see e.g,][]{2015ARA&A..53...51S, 2017ARA&A..55...59N}. The resultant connection between galaxies and dark matter haloes thus remains a topic ripe for exploration, where observational constraints can help constrain theories of galaxy formation and evolution \citep[see e.g.,][]{vdB2004, Mandelbaum:2006, 2009MNRAS.392..801M, 2010MNRAS.407....2D, 2011MNRAS.410..210M, 2015MNRAS.446.1356H, 2015ApJ...806....2M, 2018ARA&A..56..435W, Lange:2019}.

Galaxy groups lie at the crossroads of haloes with single dominant central galaxies and galaxy clusters which consist of a large number of smaller satellite galaxies. Galaxy groups are quite abundant in the Universe, and a large fraction of galaxies reside in galaxy groups \citep[][]{2004MNRAS.348..866E}. They are an important laboratory to study how different baryonic processes shape the properties of galaxies that reside in these galaxy groups \citep[][]{2010MNRAS.406..822M,2014MNRAS.441.1270L, 2020arXiv201004166G}. Scaling relations between the observable properties of galaxy groups such as their total group luminosity or group velocity dispersion with the halo mass are crucial to develop a phenomenological understanding of galaxy formation within groups \citep[see e.g.,][]{2003MNRAS.339.1057Y, 2004MNRAS.355..769E, 2007ApJ...671..153Y, 2009ApJ...703.2232S, 2011A&A...535A.105E, 2015MNRAS.446.1356H, 2015MNRAS.452.3529V,2020ApJ...892..102L, 2021Univ....7..139L, 2021MNRAS.504.4093G, 2021ApJ...909..143Y}.

The study of scaling relations requires a robust identification of galaxy groups and their corresponding observable properties. The galaxy groups consist of a smaller number of bright members, making them difficult to find in the galaxy catalogue data. We require a deep and complete galaxy catalogue sample within a large sky region to fully resolve the galaxy groups. The Galaxy And Mass Assembly (GAMA) survey \citep[][]{2009A&G....50e..12D, 2010MNRAS.404...86B} provides highly complete ($\geq 95$ percent) spectroscopic information for the galaxies in the sky region \citep[][]{2015MNRAS.452.2087L} . They use these galaxies to construct a group catalogue using a friends-of-friends (FOF) based group finding algorithm \citep[][]{1982ApJ...257..423H}. The grouping parameters are optimized by testing them on mock catalogues containing galaxies populated in dark matter-only simulations \citep[][]{2006MNRAS.370..645B}. The group catalogue provides observable properties for each group that can be used to study their connection to corresponding dark matter halo, and they also correct observables for survey selection effects \citep[][]{2011MNRAS.416.2640R}.

As the masses of groups are not directly observable, we infer them via various techniques using their observable properties. The lower richness and faintness in X-ray prevents group scale objects from obtaining reliable halo mass estimates using dynamical and X-ray techniques \citep[see e.g.,][]{2001ApJ...552..427C, Becker2007, 2011A&A...535A.105E,2020A&A...635A..36G}. Also, these methods heavily depend on the underlying assumptions - dynamical methods require a virialized system while X-ray estimation needs a hydrostatic equilibrium of intracluster gas \citep[see e.g.,][]{2013ApJ...778...74K,2015MNRAS.449.3082P,2017A&A...606A.122F}. An alternate approach for estimating halo masses for group scale objects is halo abundance matching, by measuring the abundance of galaxy groups by their group luminosity or total group stellar mass content and connecting it to the halo mass function \citep[see e.g.,][]{2007ApJ...671..153Y, 2020A&A...636A..61R,2021ApJ...909..143Y}. These estimates depend sensitively upon the assumed input cosmology.

Weak gravitational lensing is a very useful technique to get the halo masses for groups/clusters sample selected according to their observable properties \citep[for e.g.][]{2006MNRAS.372..758M, 2010PASJ...62..811O, 2014MNRAS.442.1507G, 2015MNRAS.452.3529V, 
Simet_2016, 2017MNRAS.468.3251D, 2017MNRAS.472.1946S, 2018ApJ...862....4L, 2018ApJ...854..120M, 2019MNRAS.484.1598B, 2019ApJ...875...63M, 2019PASJ...71..107M, 2020MNRAS.499.2896T, 2020MNRAS.498.5804R, 2020A&ARv..28....7U, 
2020A&A...642A..83D, 2021arXiv210305653G}. In contrast to other methods, the weak gravitational lensing gives a halo mass estimate independent of assumptions and uncertainties on the physical properties of the groups/clusters.  It maps the matter distribution in the foreground objects like groups/clusters (lens) by studying their gravitational effects on the light emitted from the background source galaxies \citep[][]{1937PhRv...51..290Z}. In weak gravitational lensing, the observed shapes of the background galaxies get distorted in a coherent pattern around the lens, dependent upon the mass distribution in the lens \citep[][]{1992ApJ...388..272K, 2001PhR...340..291B, 2008ARNPS..58...99H, 2018ARA&A..56..393M}. The individual distortions in the shapes of the galaxies are small, but these distortions can be measured statistically by stacking the weak lensing signal around many lenses together. So, to have high signal-to-noise lensing measurements, we need data from many source galaxies in the overlapping sky region. 

The Subaru Hyper Suprime-Cam (HSC) survey is a photometric survey \citep[][]{2018PASJ...70S...4A, 2018PASJ...70S...8A} that provides high-quality imaging data of the individual shapes for millions of source galaxies.  At present, HSC is one of the deepest surveys over a large region of the sky and has a high galaxy number density of $24.6\,{\rm arcmin}^{-2}$. HSC has a significant overlap with the GAMA survey footprint, making it a suitable choice for conducting gravitational lensing studies for lenses seen in the GAMA survey \citep[for strong lensing application, see][]{Holwerda2015, Chan2016}. In this work, we are using the weak lensing technique to map dark matter distribution in GAMA galaxy groups \citep[][]{2011MNRAS.416.2640R} as lens using data for sources galaxies from the HSC survey. We use group catalogue samples provided by GAMA collaboration and HSC-PDR2 S16a shape catalogues \citep[][]{2018PASJ...70S..25M, 2018MNRAS.481.3170M} for the source galaxies for the signal measurements. We also check for the systematics involved with the measurements and their effects on the final results \citep[][]{2005MNRAS.361.1287M}. In order to extract halo masses from the stacked weak lensing signals, we use a halo occupation distribution (HOD) based halo modelling scheme \citep[see e.g.,][]{2000MNRAS.318..203S, 2002PhR...372....1C, 2013MNRAS.430..725V, 2013MNRAS.430..767C,2015ApJ...806....2M} instead of a simple Navarro-Frenk-White \citep[hereafter NFW,][]{1996ApJ...462..563N} dark matter profile model fitting to the weak lensing signal measurements. This approach allows a detailed characterization of both the mean and the scatter of the halo occupation distribution of galaxy groups. 

The scaling relation between the halo mass and the group luminosity and velocity dispersion has been previously studied using the same GAMA galaxy group sample but using a weak lensing shape catalogue from the Kilo Degree Survey (KiDS) in \citet{2015MNRAS.452.3529V}. Our use of the HSC data is expected to result in improved constraints due to the factor 2 larger number density of potential sources galaxies \citep[][]{2019PASJ...71...43H}, as well as the nearly 60 percent increase in the area of overlap between the GAMA survey and the HSC compared to the KiDS. Furthermore, this is a non-trivial test of the agreement in the weak lensing signals obtained by the two independent weak lensing surveys for the same set of galaxy groups acting as lenses given the differences in shear measurement techniques, the shear calibration methods, as well as the quality of photometric redshifts, amongst others.

We describe the observational data used in Section \ref{sec:data}, the methods we used to obtain the weak gravitational lensing signal around galaxy groups in Section \ref{sec:weak_lensing} and the theoretical framework for the interpretation of the observed signal in Section \ref{sec:theory}. In Section \ref{sec:results}, we present the main results of the paper, and compare our results to those in the literature. We conclude in Section \ref{sec:conclusions} with a summary of the results and future outlook.

We perform our analysis in the context of the flat $\Lambda$CDM model with the following cosmological parameters: the matter density parameter $\Omega_{\rm m}=0.315$, the baryon density parameter $\Omega_{\rm b} h^2 = 0.02205$, the variance of density fluctuations $\sigma_8=0.829$ and the power law index of the initial power spectrum, $n_{\rm s}=0.9603$ \citep[][]{2014A&A...571A..16P}. Throughout the work, we use halo masses ($M_{\rm 200m}$) and halo boundaries ($R_{\rm 200m}$), defined to enclose a matter density equal to 200 times that of the average matter density of the Universe.

\section{Data}
\label{sec:data}
\subsection{GAMA Galaxy Group Catalog}
GAMA is a highly complete spectroscopic survey carried out using the AAOmega multi-object spectrograph on the Anglo-Australian Telescope (AAT), which spans over an area of $\sim 286\,{\rm deg}^2$ down to r < 19.8 mag. In our work, we are using full GAMA-II \citep[][]{2011MNRAS.413..971D, 2015MNRAS.452.2087L} galaxy group catalogue, which comprises three equatorial fields G09, G12, and G15 with $60\,{\rm deg}^2$ each and two (G02 and G23) southern fields over $50 \, {\rm deg}^2$ each. The galaxy groups are found by applying a FOF algorithm on galaxies from G02, G09, G12, and G15 with linking parameters optimized using mock catalogues \citep[][]{2011MNRAS.416.2640R}. These mocks are constructed from Millenium DM simulation \citep[][]{2005Natur.435..629S} by populating galaxies using a semi-analytical galaxy formation model \citep[][]{2006MNRAS.370..645B}. 
In our analysis, we use galaxy groups from the v10 of the GAMA group catalogue in its equatorial fields with at least five member galaxies in it\footnote{This selection is consistent with the criteria used in \citet{2015MNRAS.452.3529V} in order to have sufficient purity of group members and to obtain a reliable estimate of velocity dispersions \citep[see ][ for estimates of purity]{2011MNRAS.416.2640R}.} and which lie in the footprint of the HSC first year shape catalogue. We will bin our sample into group $r$-band group luminosities (as given by the \textsc{LumB} column) and velocity dispersions (as given by the \textsc{VelDisp} column) in the group catalogue. All the galaxies used for grouping are k-corrected and evolution corrected to a reference redshift of $z=0$ \citep[][]{2011MNRAS.416.2640R}. The total $r$-band luminosities are estimated using the observed group luminosity and correcting it for the fainter end using the luminosity function \citep[see eq 22 in][]{2011MNRAS.416.2640R}. Also, the velocity dispersion estimates are corrected for the velocity error for individual member galaxies in the groups.

\subsection{HSC-GAMA Survey Overlap}
\begin{figure}
    \centering
    \includegraphics[width=\columnwidth]{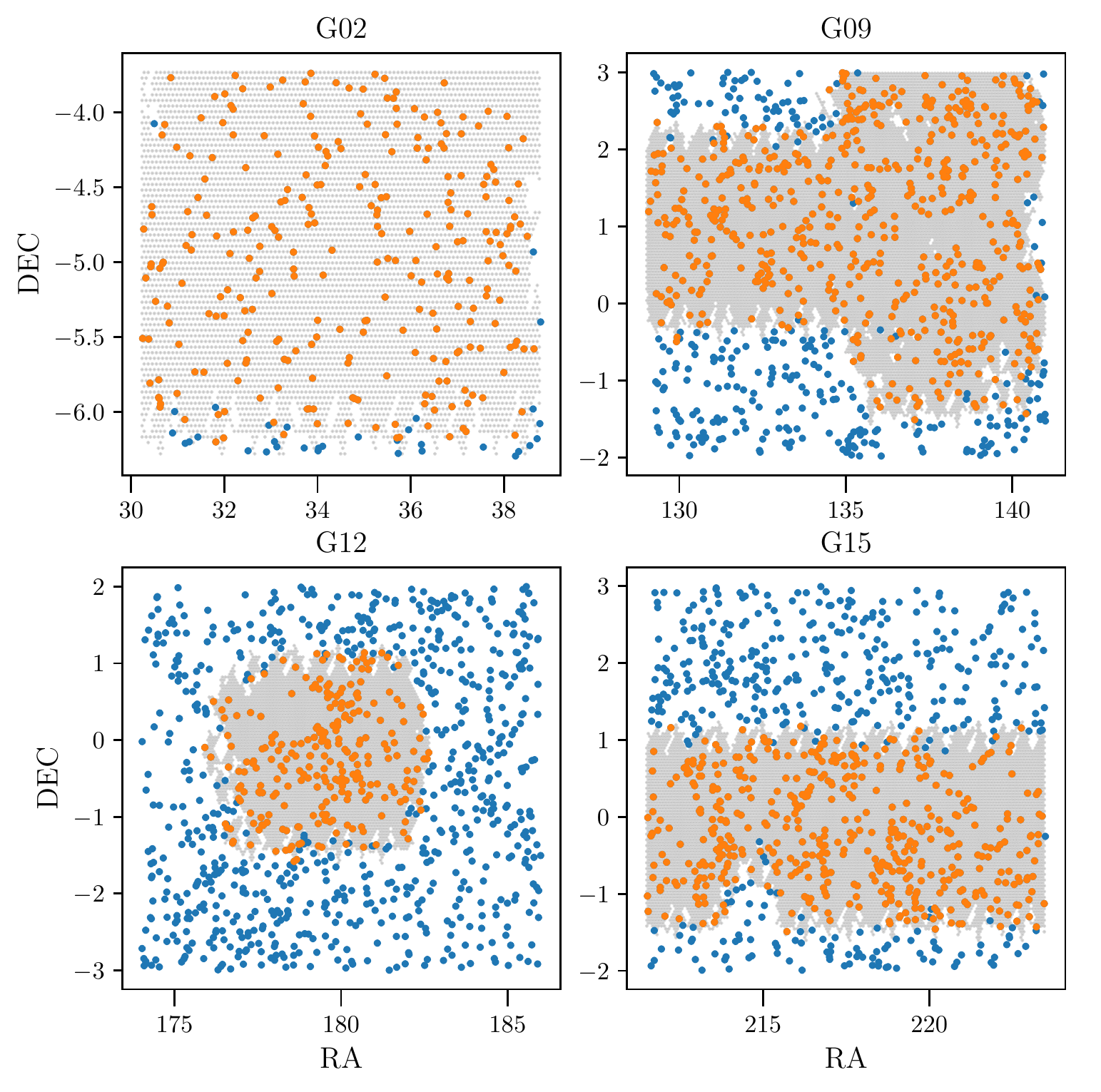}
    \caption{{\it Distribution of GAMA galaxy groups compared to the Subaru HSC footprint on the sky}: The points in the different panels of the figure show the distribution of GAMA galaxy groups with a minimum of 5 members in the GAMA02H, GAMA09H, GAMA12H and GAMA15H fields, respectively. The grey shaded region indicates the area which overlaps with the HSC year 1 shape catalogue data from data release S16a. The orange points correspond to galaxy groups within the HSC footprint that we use in our analysis, while the blue points lie outside the footprint. The GAMA survey has an unmasked area overlap of 99.9 deg$^2$ with the first year shape catalogue from the HSC survey.}
    \label{fig:skymap}
\end{figure}

The GAMA group catalogue has an overlap of over $103\,{\rm deg}^2$ with the footprint of the first year shape catalogue of the HSC survey. Once the star mask is accounted for, this overlap area reduces to  $\sim 99.9 \,{\rm deg}^2$. This results in a total of 1587 galaxy groups which we use for our analysis. We apply selection cuts on group properties to study their correlation with group halo masses. We have selected groups with at least five members and binned them in group luminosity and velocity dispersion. Figure \ref{fig:skymap} shows galaxy groups from the four GAMA fields (G02, G09, G12, G15) within the HSC S16a footprint. The orange points indicate the groups which lie within the HSC S16a footprint (shown with gray shaded region), while the blue points indicate the groups that lie outside the footprint. The galaxy groups span over the redshift range of $0<z<0.5$ with a median redshift of 0.2. We do not keep any buffer between the GAMA groups and any survey edges or mask boundaries in the HSC data. Given the depth of each exposure in the HSC survey, the portion of the HSC area that is covered by star masks is not insignificant. Thus avoiding every boundary would make this approach impractical.

\begin{table*}
\begin{tabular}{ |p{2cm}|p{4.5cm}|p{4cm}|p{4cm}|}
 \multicolumn{4}{|c|}{} \\
 \hline
 Observable & Selection cuts & Number of lenses & Mean redshift\\
 \hline
 $\log[L_{\rm grp}/ (h^{-2} L_\odot)]$ & (9.4, 10.9, 11.1, 11.3, 11.5, 11.7, 12.7) & (388, 279, 316, 255, 184, 165) & (0.12, 0.18, 0.21, 0.26, 0.28, 0.33)\\
 $\sigma / (\kms)$ & (0, 225, 325, 375, 466, 610, 1500) & (488, 397, 166, 216, 173, 141) & (0.16, 0.20, 0.21, 0.23, 0.26, 0.29) \\
 \hline
\end{tabular}
\caption{ The table shows the bin edges we use to divide GAMA galaxy groups into six bins each of group luminosity and velocity dispersion, respectively. We indicate the number of lenses and mean redshifts for galaxy groups with at least five members. We identify the brightest cluster/group galaxy (BCG) as the group centre. Here we consider only those groups which lie within the HSC S16a footprint.}
\label{table_1}
\end{table*}

In Table \ref{table_1} we provide the selection cuts used on the group observables - group luminosities and velocity dispersion, with the number of groups/lenses in each bin and the corresponding mean redshifts. We have divided our sample into six observable bins, and we compute the weak lensing signal for the lenses in each bin, which we model to get a constraint on the mean halo masses of these groups. In our analysis, we use the brightest cluster/group galaxy (BCG) as the group centre for the weak lensing signal computations and to obtain the mean halo mass constraints. We have also verified our results by using other group centres, and they are in good agreement with each other.

For checks on weak lensing systematics, we use 32 times more random points than the number of GAMA galaxy group lenses in each bin. The randoms are distributed uniformly on the sky within the HSC-GAMA overlap area. Their redshifts are drawn randomly, with replacement, from the redshifts of the lens sample and therefore, statistically, they follow the same redshift distribution as our lens sample. 

\subsection{HSC Shape and Photo-$z$ Catalogs}
The Hyper Suprime Cam is a large field of view camera ($1.77 \,{\rm deg}^2$) situated at the prime focus of the 8.2m Subaru telescope \citep[][]{2018PASJ...70S...2K,2018PASJ...70S...1M}
located on Maunakea in Hawaii. When combined with the excellent on-site seeing conditions (median seeing of $\sim 0.6"$), the HSC instrument is suited for a large scale weak lensing survey. In year 2014 under the Subaru Strategic Program (SSP) \citep[][]{2018PASJ...70S...4A}, the HSC survey collaboration started an imaging survey to observe $1400 \,{\rm deg}^2$ of the sky with an unprecedented depth of $i=26$ by the end of the fifth year of operations. The first year shape catalogue \citep[][]{2018PASJ...70S..25M} is based on the data taken from April 2014 to April 2016 and corresponds to about 90 nights of observations in total. The first-year data cover over $136.9 \,{\rm deg}^2$ of the area in six different fields - HECTOMAP, GAMA09H, WIDE12H, GAMA15H, XMM, and VVDS. In our work, we use the shape catalogue from data release S16a \citep[][]{2019PASJ...71..114A}, which is a slight extension of the first public data release from HSC Survey \citep[][]{2018PASJ...70S...8A}, but has been made available as an incremental data release. The shape catalogue has an effective galaxy number density of $21.5\,{\rm arcmin}^{-2}$ at a median redshift of $0.8$. In our analysis, we will use data from HSC fields - GAMA09H, WIDE12H, GAMA15H, XMM as they overlap with the GAMA group catalogue sky region.

The shapes of the galaxies were estimated by applying re-Gaussianization \citep[][]{2003MNRAS.343..459H} PSF correction technique on the coadded i-band images. This method has been used and well studied for the data from the SDSS survey \citep[][]{2005MNRAS.361.1287M,2012MNRAS.425.2610R,2013MNRAS.432.1544M}. It provides shapes $(e_1, e_2) = (e\cos 2\phi, e\sin 2\phi)$ where $e = (a^2 - b^2)/(a^2+b^2)$ where $a$ denotes the semi-major and $b$ the semi-minor axis of the galaxies \citep[][]{2002AJ....123..583B} and $\phi$ denotes the position angle of the major axis with respect to the equatorial coordinate system. These shape estimates are further calibrated with image simulations produced via \textsc{GALSIM} \citep[][]{2015A&C....10..121R} - an open source software package, but which mimic the observing conditions of the HSC survey \citep{2018MNRAS.481.3170M}. These image simulations are used to estimate the additive biases $(c_1,c_2)$, the multiplicative bias $m$ of the shear estimation, the rms ellipticity $e_{\rm rms}$ of the intrinsic shapes of the galaxies and the shape measurement error $\sigma_{e}$ for every galaxy. The rms ellipticity and measurements are then used to assign minimum variance weights $w_s = (e^2_{\rm rms}+\sigma_{e}^2)^{-1}$ for each galaxies  \citep[for more details see][]{2018MNRAS.481.3170M}. Further cuts are applied on the shape catalogue data for weak lensing cosmology as described in  \citep[][]{2018MNRAS.481.3170M}.

For each of the galaxies in the shape catalogue, HSC-SSP provides photometric redshift estimates using six different methods \citep[][]{2018PASJ...70S...9T}. In our work, we use the full redshift PDF $P(z)$ for the galaxies computed by running classical template fitting code \textsc{Mizuki} \citep[][]{2015ApJ...801...20T}. We also apply selection cuts on $P(z)$ to filter galaxies for the analysis as described in section \ref{sec_3.1}.

\section{Weak gravitational lensing}
\label{sec:weak_lensing}

\subsection{Stacked ESD Profile}
\label{sec_3.1}
The weak gravitational lensing signal at a galaxy group centric comoving radial distance $R$ is given by
\begin{equation}
    \label{s3eq1}
    \Delta\Sigma(R) = \overline{\Sigma}(<R) - \langle \Sigma(R)\rangle = \Sigma_{\rm crit}\langle \gamma_t \rangle
\end{equation}
where $\Delta\Sigma(R)$ is the excess surface density (ESD), $\overline{\Sigma}(< R) = \left(\int_0^R \Sigma(R') \, 2 R' dR'\right)/ R^2$ denotes the average surface density of mass within a given projected  radius $R$, while $\langle \Sigma(R) \rangle$ denotes the azimuthally averaged projected surface density at radius R and $\langle \gamma_t \rangle$ denotes the average tangential shear and $\Sigma_{\rm crit}$ is the critical surface density which depends on the redshifts of the source and the lens and is given by
\begin{equation}
    \Sigma_{\rm crit} = \frac{c^2}{4\pi G }\frac{D_{\rm a}(z_{s})}{(1+z_{l})^2D_{\rm a}(z_{l})D_{\rm a}(z_{l},z_{s})}\,.
\end{equation}
Here $D_{\rm a}(z_{l})$, $D_{\rm a}(z_{s})$ and $D_{\rm a}(z_{ l},z_{s})$ denote the angular diameter distances for the lens at redshift $z_{l}$, the source at redshift $z_{s}$, and the lens-source pair, respectively. The factor of $(1+z_{l})^2$ in the denominator converts the physical critical density to comoving coordinates. The inverse critical density $\Sigma_{\rm crit}^{-1} = 0$ for the lens source pairs with $z_s \leq z_l$.

For computing the weak lensing signal for our lensing sample, we follow the methodology described in appendix A.3.1 of the first year HSC shape catalogue paper \citep[see][]{2018PASJ...70S..25M}. The shape catalog provides measurement of the distortions ($e_1$, $e_2$) for each source galaxy along with its corresponding shape weight $w_{s}$. The catalog also provides additive $(c_1, c_2)$ and multiplicative $m_{s}$ biases for each background galaxy calibrated using detailed image simulations ran for HSC-like conditions \citep[][]{2018MNRAS.481.3170M}. From the data, we compute the ESD $\Delta\Sigma(R_i)$ at each comoving radial bin $R_i$ as
\begin{equation}
\Delta\Sigma(R_i) =  \frac{1}{(1+\hat{m})}\left(\frac{\sum_{ ls} w_{ls} e_{t,ls}\langle \Sigma_{\rm crit}^{-1}\rangle^{-1}}{2 \mathcal{R}\sum_{ls} w_{ls}}-\frac{\sum_{ls} w_{ls} c_{t,ls}\langle \Sigma_{\rm crit}^{-1}\rangle^{-1}}{\sum_{ls}w_{ls}} \right)
\end{equation}
where the summation is over all sources-lens pairs in a given radial bin $R_i$. The $e_{\rm t,ls}$ and $c_{\rm t,ls}$ are tangential components of ellipticies and additive bias with a weight $w_{\rm ls} = w_s \langle \Sigma^{-1}_{\rm crit}\rangle^2$ for each source-lens pairs. Given that we have a probability distribution for the source redshift $p(z_{\rm s})$, we define $\langle \Sigma^{-1}_{\rm crit}\rangle$ averaged over $p(z_{\rm s})$, such that 
\begin{equation}
\langle\Sigma^{-1}_{\rm crit}\rangle = \frac{4\pi G(1+z_{ l})^2}{c^2}\int_{z_{l}}^{\infty}\frac{D_{\rm a}(z_{l}) D_{\rm a}(z_{l},z_{s})}{D_{\rm a}(z_s)} p(z_s)\, dz_s \,.
\end{equation}
In our study, we only use sources with $\int_{z_{\rm min}}^{\infty} p(z_s)\,dz_s > 0.99$, where $z_{\rm min} = z_{\rm lmax} + z_{\rm diff}$. We have used $z_{\rm lmax} = 0.472$ the maximum redshift of our lens sample and $z_{\rm diff} = 0.1$. These cuts ensure a secure selection of galaxies which lie in the background of the lens galaxies and thus mitigate any contamination from correlated galaxies at the lens redshift which could potentially dilute the weak lensing signal. After application of these cuts we are left with an effective galaxy number density of $11.95\,{\rm arcmin}^{-2}$.  The quantity $\mathcal{R}$ denotes the shear responsivity which represents how the measured ellipticities $e_1$ and $e_2$ respond to small values of shear and can be computed using the per object RMS distortions $e^2_{\rm rms} $\citep[][]{2002AJ....123..583B}, such that,
\begin{equation}
    \mathcal{R} = 1-\frac{\sum_{ls} w_{\rm ls} e^2_{{\rm  rms},ls}}{\sum_{\rm ls} w_{ls}}\,. 
\end{equation}
We have also applied multiplicative $\hat{m}$ calibration biases given as an ensemble average over source-lens pairs
\begin{equation}
\hat{m}=\frac{\sum_{ls} m_s w_{ls}}{\sum_{ls} w_{ ls}}    
\end{equation}
Apart from the above mentioned biases we also apply a multiplicative bias ($m_{\rm sel}$) due to a lower cut adopted on the resolution factor ($R_2 \geq 0.3$) in the weak lensing catalog \citep[see ][]{2018MNRAS.481.3170M}). This selection bias is given by $m_{\rm sel} = A p(R_2 = 0.3)$ with $A=0.00865$, where $p(R_2=0.3)$ denotes the probability density of galaxies at the edge of the sample. This probability $p(R_2 = 0.3)$ is computed after accounting for the source-lens weights $w_{\rm ls}$ for each radial bin $R_i$. 
\begin{figure}
    \centering
    \includegraphics{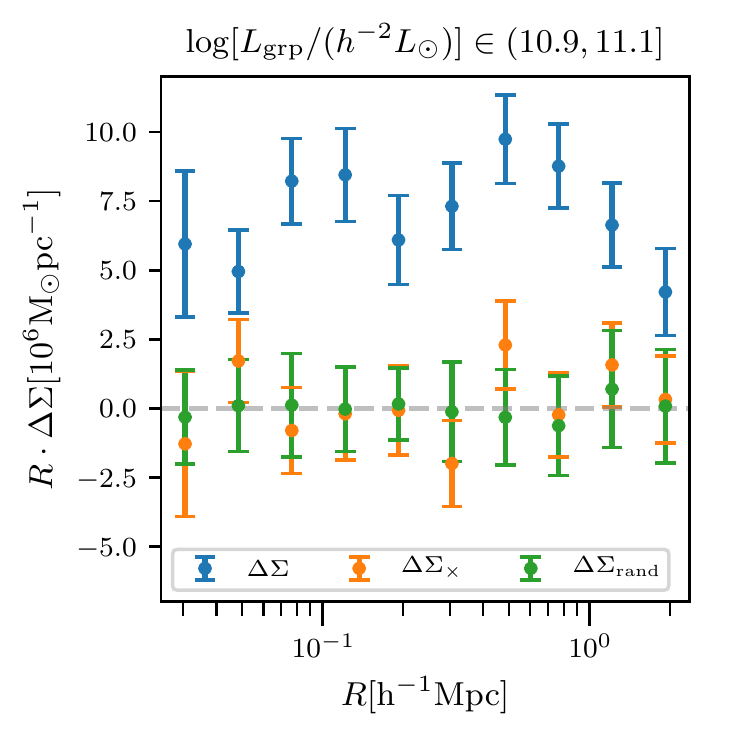}
    \caption{{\it Systematics tests for the weak lensing measurements}: The blue points with errors show measured values of $\Delta \Sigma$ for the group luminosity bin $\log L_{\rm grp}\in[10.9, 11.1]$ along with shape noise errors. The orange data points represent the cross-component $\Delta \Sigma_{\times}$ for the same bin with shape noise errors and are consistent with zero. The green points with errors show the value of $\Delta\Sigma$ measured around random points having the same redshift and on sky distribution as the galaxy groups in this selection bin. The errors on the green points correspond to $1\sigma$ standard deviation from 32 random realizations. This signal around random points has been already subtracted from the blue points. This bin is representative of the results that we get for other bins and selections.}
    \label{fig:null}
\end{figure}

Even though we use stringent cuts on the photometric redshifts of galaxies, we estimate the possible dilution of the signal due to the use of source galaxies correlated with the lensing galaxies. The resultant dilution of the signal due to unlensed galaxies can be studied using a boost factor $C(R_i)$ \citep[][]{2004MNRAS.353..529H, 2005MNRAS.361.1287M, 2013MNRAS.432.1544M, 2015ApJ...806....1M, 2018ApJ...854..120M} which is a ratio between the weighted number of source-lens pairs for lensing sample to the randoms for a given radial bin $R_i$ and  defined as:
\begin{equation}
    C(R_i) = \frac{N_r \, \sum_{ls} w_{ls}}{N_l \, \sum_{ rs} w_{ rs}}\,.
    \label{eqn_7}
\end{equation}
Here $N_l$ and $N_r$ are the number of lenses and randoms used for the computation of the signal, and $w_{\rm rs}$ is the random-source weight similar to $w_{\rm ls}$ lens-source weight. The randoms are expected to have the same redshift distribution as that of the lensing sample but with random positions in the sky. These random points are expected to obey the same survey geometry and masks as present in the real data. The boost factor $C(R_i)$ needs to be multiplied to the ESD profile to correct the dilution of the signal. In Appendix \ref{app:boost} we present the boost parameter analysis for each selection bins in our lensing sample.
 
We also study the systematic bias in the photometric redshift estimates of the source galaxies which can affect the ESD ($\Delta\Sigma$) measurements via the computation of the critical surface density ($\Sigma_{\rm crit}$). We estimate this bias using eqn 5 given in \citet[][]{2008MNRAS.386..781M} for a lens sample at redshift $z_l$ as
 \begin{equation}
     \frac{\Delta \Sigma}{ \widetilde{\Delta\Sigma} }(z_l) = 1 + b(z_l) = \frac{\sum_s w_{ls}\langle\Sigma^{-1}_{{\rm crit},ls}\rangle \widetilde{\Sigma}^{-1}_{{\rm crit},ls}}{\sum_s w_{ls}}\,.
     \label{eqn_8}
 \end{equation}
The quantities with tilde represent computations done using the true redshifts of source galaxies, and we are doing a summation over all the source galaxies. Given the depth of our source galaxy catalogue, it is difficult to have a spectroscopic sample that mimics the same population properties. Instead, we use the COSMOS-30 band photo-$z$ sample \citep[][]{2009ApJ...690.1236I} with a weighting $w_{\rm SOM}$ to calibrate the COSMOS-30 band photo-$z$ galaxies to match the colour and magnitude distribution of the source galaxy sample. This weighting is provided along with the publicly released weak lensing shape catalogue by the HSC survey. We include these weights $w_{\rm SOM}$ in $w_{\rm ls}$ while doing the computations. This method for photo-$z$ bias computations has been used in various studies in the past \citep[e.g.][]{2012MNRAS.420.3240N, 2019ApJ...875...63M, 2019PASJ...71..107M}. We checked for the photo-$z$ bias using Eqn \ref{eqn_8} as a function of lens redshift $z_l$ and computed an average photo-$z$ bias for our lensing sample by applying suitable weights to the lenses as described in \citep[see Eqn 23 in][]{2012MNRAS.420.3240N}. The average value of the bias parameter over the redshift range of our lensing sample is about one percent which is negligible.  

\begin{figure*}
    \centering
    \includegraphics[width=\columnwidth]{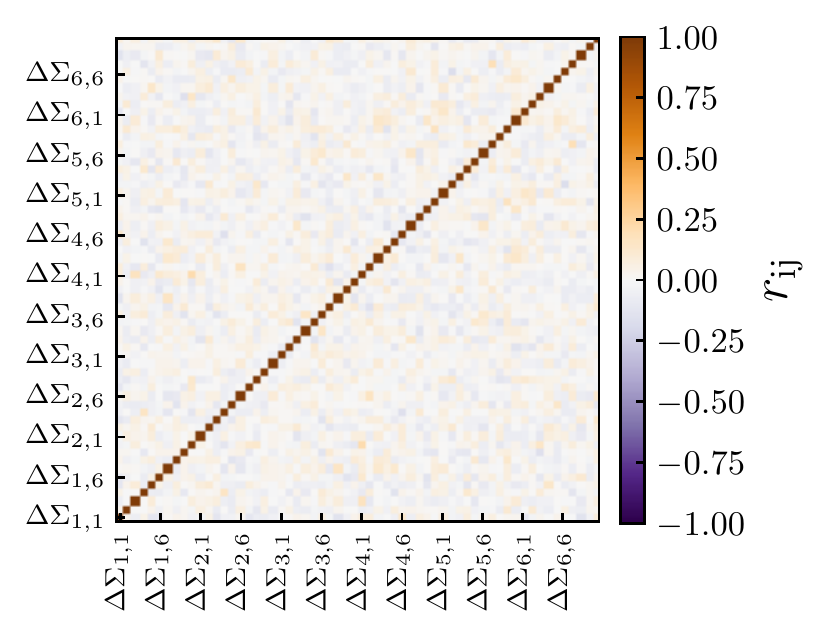}
    \includegraphics[width=\columnwidth]{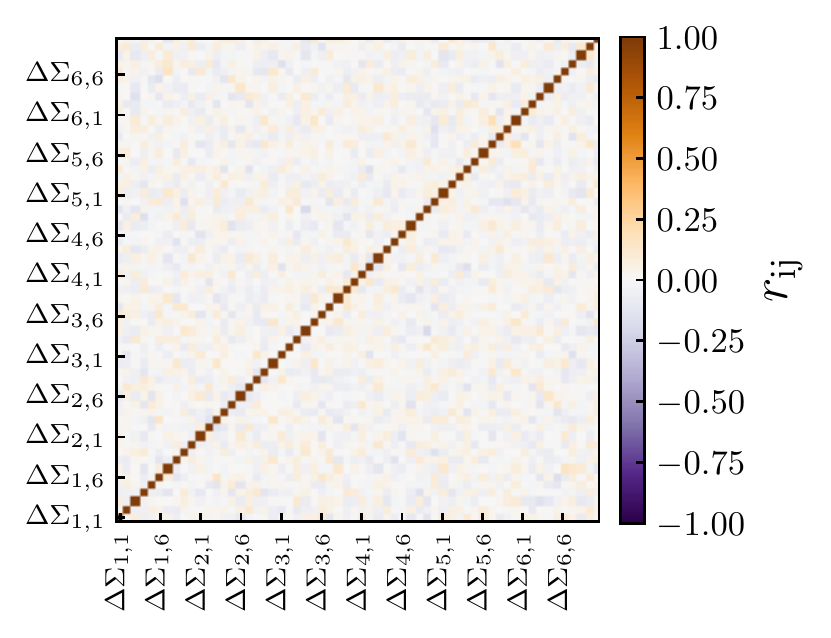}
    \caption{{\it Covariance measurements}: We obtain the shape noise covariance matrix by randomly rotating the shapes of galaxies in the HSC shape catalogue 320 times and measuring the resultant signal around GAMA galaxy groups with at least five members and resides within HSC S16A footprint. The left and right panels of the figure show the correlation coefficient $r_{ij}$ of the errors on our measurements of galaxy groups binned by group luminosity and group velocity dispersion, respectively. The quantity $\Delta \Sigma_{i,j}$ in the figure represents the $j^{\rm th}$ radial bin of the $i^{\rm th}$ selection bin. We observe very little covariance between the measurements in the different bins.}
    \label{fig:cov}
\end{figure*}

After considering the biases mentioned above, we also compute ESD signals around the random points and subtract them from the signals around the lensing sample. It helps us to measure the ESD profile for lenses over the background and further correct for any additive biases in the shear estimations due to PSF corrections \citep[for more details see][]{2004AJ....127.2544S, 2005MNRAS.361.1287M, 2017MNRAS.471.3827S}.

In Figure \ref{fig:null}, we show the ESD signal measurements for a $r$-band group luminosity selection bin ($\log[L_{\rm grp} / (h^{-2} L_\odot)] \in (10.9,11.1]$). The blue data points represent the ESD profile $\Delta\Sigma$ with signals around random points subtracted from it along with shape noise error bars. Given that the shear field around lenses should not possess any handedness, we expect the cross-component to be consistent with zero. We present these null tests for our signal measurements by computing the ESD signals for the cross-component, $\Delta \Sigma_{\times}$, and around random points, $\Delta \Sigma_{\rm rand}$, in the same selection bin. The cross ESD signal $\Delta \Sigma_{\times}$ is shown by orange points along with shape noise errors, and the green data points are measurements around the random points with errors from the scatter in random measurements. We see that both the signals are consistent with the null detection within the given uncertainties, and we see similar results for the other selection bins. 

\subsection{Covariance}
\label{sec_3.2}
The same source galaxy in the shape catalogue can contribute to the signal for multiple lens groups at different radial bins. This can create a covariance between the measurements of the stacked ESD profile at a different radial distance away from the group centre. In order to estimate this covariance, we randomly rotate the source galaxies around their positions and then use these rotated ellipticities to compute the ESD signal. The random rotations will erase the coherent tangential shear pattern. The signal measured using the shapes of these randomly rotated galaxies can be used to estimate the intrinsic shape noise of the source galaxies. We rotated source galaxies 320 times for each sample selection bin in table \ref{table_1} and computed the ESD signal for each case. As the positions of the source galaxies are preserved in this procedure, the covariance in the different radial bins, due to the use of the ellipticity of the same source galaxy, is preserved. We use these measurements to compute the covariance among different radial bins for various selection cuts. In weak lensing, the coherent distortions are tiny compared to the intrinsic shapes of the galaxies; therefore, the shape noise dominates the errors on the ESD signal measurements, especially on small scales. Therefore, we use error bars given by the shape noise for the ESD profile for our lensing sample. This is similar to the methodology adopted in the study of the GAMA groups using source galaxies from the KiDS survey \citep[][]{2015MNRAS.452.3529V}. We compute the cross-correlation coefficient $r_{ij}$ defined as
\begin{equation}
    r_{ij} = \frac{C_{ij}}{\sqrt{C_{ii} C_{jj}}}
    \label{corr}
\end{equation}
where $C_{ij}$ is the component of covariance matrix $\textbf{C}$ determined using shape noise for each of the lens selection cuts. In Figure \ref{fig:cov},
we showed the correlation coefficient for both group luminosity and velocity dispersion selection cuts. We labelled both the $x$ and $y$-axes as $\Delta \Sigma_{i,j}$ which refers to the $j^{\rm th}$ radial bin of $i^{\rm th}$ group selection bin. The off-diagonal terms of the cross-correlation coefficient are quite negligible, and we observe fairly uncorrelated measurements for both the radial positions within a given selection bin as well as those between different selection bins. We primarily use the covariances from shape noise for the purpose of modelling the ESD profile and the square root of this covariance matrix as shape noise error bars. We also check for possible effects of inclusion of the large scale structure variance on our results. We have recomputed the covariances using the jackknife technique \citep[][]{1982jbor.book.....E,2009MNRAS.396...19N} with 100 jackknife regions with an area of about $1.0\,{\rm deg}^2$ each. Even with the jackknife technique, we do not see any significant evidence for covariances between the different bins used in our analysis. We also compare the results of our analysis using the shape noise covariance and the jackknife covariance in Appendix \ref{app:jackknife}, and show that the resultant constraints do not differ significantly from each other.

\section{Halo Model}
\label{sec:theory}
\label{sec_4}
In this section, we describe the theoretical framework used for the modelling of the ESD measurements. The halo model \citep[][]{2000MNRAS.318..203S, 2002PhR...372....1C, 2013MNRAS.430..725V} provides a statistical description of the measurements and allows us to the interpretation of the results. The ESD signal depends on the projected matter density profile $\Sigma (R)$, and we can express it in terms of the cross-correlation function $\xi_{\rm gm} (r)$ of matter with the central group galaxy which acts as a baryonic tracer for the centre of the dark matter haloes. This projected surface density of matter around the group centre is given
\begin{equation}
    \Sigma (R) = \bar \rho_{\rm m} \int_{-\infty}^{\infty} \left(1+\xi_{\rm gm} (\sqrt{R^2 + \pi^2})\right) \, d\pi
\end{equation}
where $\pi$ denotes the distance in the line-of-sight direction and $R$ denotes the halo-centric projected radial distance. We use the current mean matter density $\bar \rho_{\rm m}$ as we are computing projected densities in comoving coordinates. The cross-correlation function $\xi_{\rm gm} (r)$ can be expressed in terms of the halo mass function ${ n(M)}$, the halo occupation distribution (HOD) ${ P(X_{\rm grp}|M)}$ which is the probability that a group satisfying our selection function ($X_{\rm grp}$) resides in a halo mass $M$ and halo-matter correlation function ${ \xi_{\rm hm} (r;M)}$ for haloes of mass $M$.
\begin{equation}
    \xi_{\rm gm} (r| X_{\rm grp}) = \frac{1}{\Bar{n}_g} \int dM \, P(X_{\rm grp}|M) \, n(M)\, \xi_{\rm hm}(r;M)\,.
\end{equation}
The denominator $\Bar{n}_g = \int dM \,P(X_{\rm grp}|M)\, n(M)$ corresponds to the mean number density of galaxy groups. The halo-matter cross correlation $\xi_{\rm hm} (r;M)$ can itself be written as a Fourier transform of the halo-matter cross spectra $P_{\rm hm} (k;M)$ such that
\begin{equation}
    \xi_{\rm hm} (r;M) = \int^{\infty}_{0} \frac{k^2 dk}{2\pi^2} P_{\rm hm}(k;M) j_0(kr)\,, 
\end{equation}
where $j_0(kr)$ is the zeroth order spherical Bessel function. The ESD $\Delta\Sigma(R;M)$ for a halo of mass M can be written using the second order Bessel function of the first kind, $J_2(kR)$ and halo-matter cross spectra $P_{\rm hm}(k;M)$ \citep[see e.g.,][]{2018ApJ...854..120M}.
\begin{equation}
    \Delta\Sigma(R;M) = \bar \rho_{\rm m} \int^{\infty}_{0} \frac{k dk}{2\pi} P_{\rm hm}(k;M) J_2(kR) 
\end{equation}
We assume that the HOD $P(X_{\rm grp}|M)$, which corresponds to the fraction of haloes of a mass $M$ that hosts galaxies in a particular bin in group observable, is proportional to a log-normal distribution following \citet{2015MNRAS.452.3529V} who use a similar lensing sample,
\begin{equation}
    P(X_{\rm grp}|M) \propto \frac{1}{\sqrt{2\pi} \sigma_{\log M'}[X_{\rm grp}]} \exp\left[-\frac{\left(\log M - \log M'[X_{\rm grp}]\right)^2}{2 \sigma^2_{\log M'}[X_{\rm grp}]}\right]
\end{equation}
where $\log M'$ and $\sigma_{\log M'}$ correspond to the mean and the spread in the halo masses for a given lens sample\footnote{As indicated in \citet{2015MNRAS.452.3529V}, this functional form should not be given a larger physical meaning than a distribution that characterizes the occupation distribution of the GAMA galaxy groups.}.

The adopted centre of the dark matter halo might differ from the true centre of the halo and, if not appropriately addressed, can lead to a biased measurement of halo masses \citep[see e.g.,][]{2012ApJ...757....2G, 2020MNRAS.493.1120Y}. We model this mis-centring in a statistical manner \citep[see][]{2011PhRvD..83b3008O, 2015ApJ...806....2M} by splitting the contribution into the fraction of off-centred cases $f_{\rm off}$ and assume that this mis-centring follows a 3-dimensional Gaussian distribution with width $r_{\rm off}$ (in units of $R_{200m}$) expressed as
\begin{equation}
    \mathcal{H}_{\rm off}(k;M) = \left(1 - f_{\rm off} + f_{\rm off} \exp\left[-\frac{k^2}{2}(R_{200m}r_{\rm off})^2\right]\right)\, P_{\rm hm}(k;M)
\end{equation}
This implies that the ESD $\Delta\Sigma(R; M)$ for a halo of mass M can be written as
\begin{align}
\Delta\Sigma(R;M)&= \bar \rho_{\rm m} \int^{\infty}_{0} \frac{k dk}{2\pi} \mathcal{H}_{\rm off}(k;M) J_2(kR)\,
\end{align}
and the total ESD for the selected galaxy groups is given by
\begin{align}
\Delta\Sigma(R) &=\frac{1}{\bar{n}_g} \int dM n(M) P(X_{\rm grp}|M) \Delta\Sigma(R;M)\,.
\end{align}
In our work, we use theoretical predictions from the \textsc{Dark Emulator}, a cosmological N-body simulation-based emulator which predicts the statistical halo properties such as the halo mass function, the halo-matter cross-correlation function, and the halo-halo correlations as a function of halo masses for a given cosmology in a redshift range of 0 to 1.48. These quantities can then be combined to predict the ESD measurements as a function of comoving radial bins \citep[see][for more details]{2019ApJ...884...29N, 2021arXiv210100113M}. Our modeling scheme differs from the analytical approach for NFW profile-based HOD modeling used in the earlier study by \citet[][]{2015MNRAS.452.3529V}. We infer the mean halo mass $\langle M \rangle$ for our groups by using the equation 
\begin{align}
\langle M \rangle = \int \mathcal{P}(M|X_{\rm grp}) \,M dM\label{eqn_19}
\end{align}
where $\mathcal{P}(M|X_{\rm grp})$ is the probability that a group with our selection resides in a halo of mass $M$. This probability can be obtained from $P(X_{\rm grp}| M)$ using the Bayes' theorem,
\begin{align}
\mathcal{P}(M|X_{\rm grp})&= \frac{ P(X_{\rm grp}|M) \, n(M)}{\bar{n}_{\rm g}}\label{eq:avgmass}\,.
\end{align}
Apart from the dark matter component, we also model the baryonic contribution to the ESD measurements. In principle, we can do full modelling of the baryonic component itself by assuming baryonic distribution profiles \citep[e.g., ][]{2015MNRAS.449.2128K} but for the scales of interest in our analysis, we model this contribution as a point mass ($M^*$) contribution to the ESD measurements ($\Delta\Sigma_b(R)$) such that
\begin{equation}
    \Delta\Sigma_b(R) = \frac{M_*}{\pi R^2}
\end{equation}
We add the baryonic contribution to the ESD profile in order to predict the total ESD signal which can be used for inference from our measurements. Our five parameter model comprises of $\Theta$ = ($\log M', \sigma_{\log M'}, f_{\rm off}, r_{\rm off}, \log M_*$). We carry out a Bayesian analysis to infer the posterior distribution of our parameters given the data. We assume fairly uninformative priors for each of our parameters in the analysis, and these priors are listed in Table \ref{table_2}. The posterior distribution of our model parameters given the data is given by the Bayes theorem,
\begin{align}
    P(\Theta|\mathcal{D}) &\propto P(\mathcal{D}|\Theta) P(\Theta)
\end{align}
where $P(\mathcal{D}|\Theta)$ is the likelihood of the data $\mathcal{D}$ given the model parameters $\Theta$, and $P(\Theta)$ corresponds to the prior probability distribution of the parameters. We assume that the likelihood to be a Gaussian,
\begin{equation}
    P(\Theta|\mathcal{D}) \propto \exp\left[-\frac{\chi^2(\Theta)}{2}\right] P(\Theta)\,,
\end{equation}
where
\begin{equation}
    \chi^2(\Theta) = \sum [\Delta\Sigma_{\rm mod}^i - \Delta \Sigma^i]^{T}\, C^{-1}_{ij} \, [\Delta \Sigma_{\rm mod}^j - \Delta \Sigma^j]\,.
\end{equation}
Here, $\Delta \Sigma^i$ is the ESD measurement and $\Delta\Sigma_{\rm mod}^i$ is the model prediction given the parameters $\Theta$ in $i^{\rm th}$ radial bin and $C^{-1}$ is the inverse of the covariance matrix $C$. We use Markov Chain Monte Carlo (MCMC) based package \textsc{emcee} \citep[][]{2013PASP..125..306F} for sampling the posterior probability distribution for model parameters. We ran chains using 256 walkers with 1000 steps for the burn-in phase and 3000 steps for the model evaluations.

\begin{table}
    \centering
    \begin{tabular}{|p{3cm}||p{3cm}|}
    \hline
    \multicolumn{2}{|c|}{Model Parameters} \\
    \hline
    Parameters& Priors \\
    \hline
    ${\rm \log [M^\prime/h^{-1} M_\odot]}$   & flat [12,15.9] \\
    $\sigma_{\rm \log M'}$&   flat [0.05,1.5] \\
    $f_{\rm off}$ & flat [0,1] \\
    $r_{\rm off}$ & flat [0,0.5]  \\
    ${\rm \log [M_* /h^{-1} M_\odot]}$  &  flat [10,12.5] \\
    \hline
    \end{tabular}
    \caption{The table shows the prior distributions of the parameters used for the modeling of the ESD measurements for GAMA galaxy groups having at least five members. The parameters ${\rm \log M'},\sigma_{\rm \log M'}$ describe the halo occupation parameters, the parameters $f_{\rm off}, r_{\rm off}$ corresponds to our nuisance parameters related to off-centering, while ${\rm \log M_*}$ is a parameter that captures the baryonic contribution within the BCG.}
    \label{table_2}
\end{table}

\begin{figure*}
    \centering
    \includegraphics[width=\textwidth]{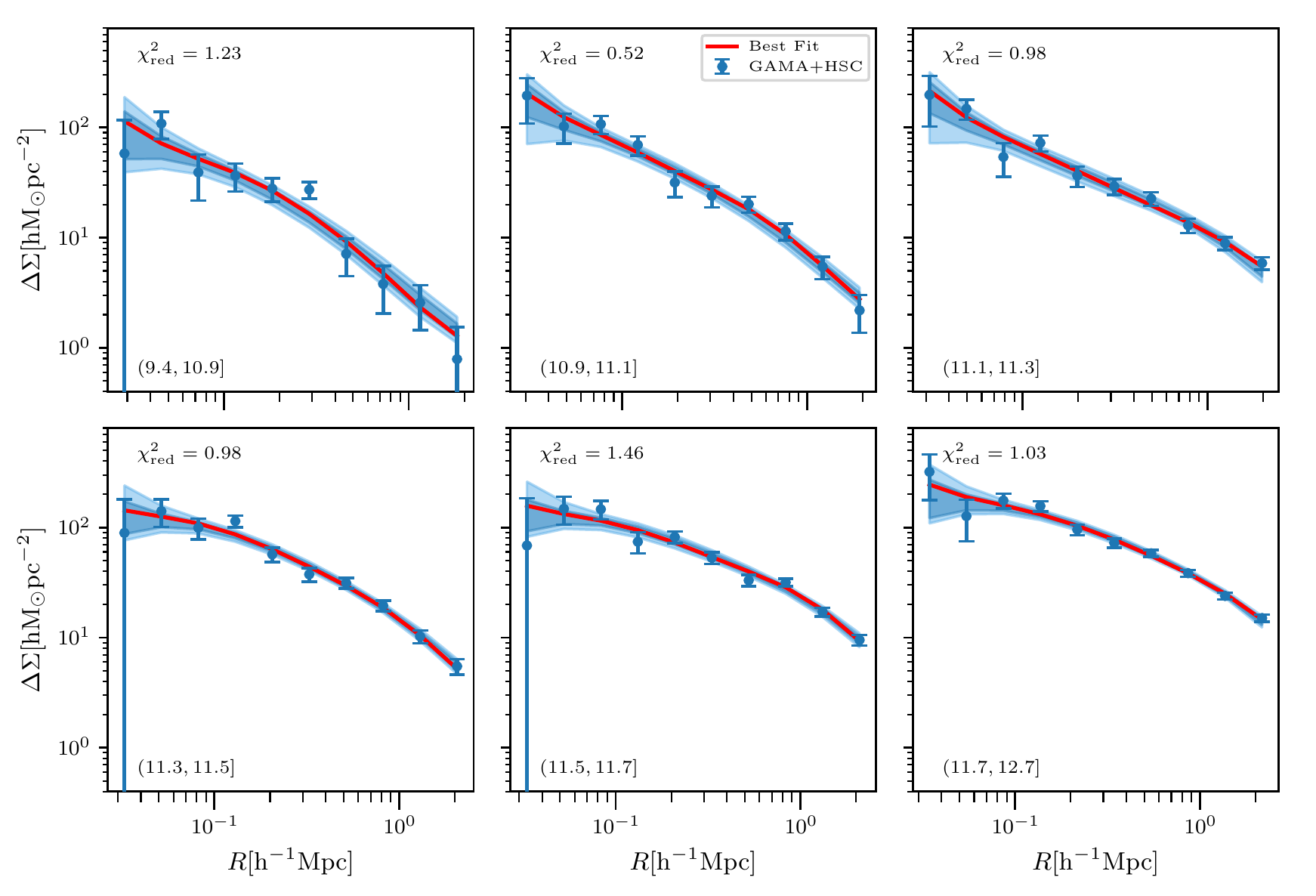}
    \caption{{\it Model fits:} The blue points with errors correspond to our measurements of the weak lensing signal for galaxy groups binned in group luminosity (see Table \ref{table_1}). The red line in each panel shows the best fit model prediction, and the corresponding $\chi^2_{\rm red}$ values are indicated in the top right hand of each panel. The dark and light blue shaded regions are the 68 and 95 percentile predictions of our constrained model given the weak lensing measurements. The corresponding best fit parameters can be found in Table \ref{table_3}.} 
    \label{fig:modlum}
\end{figure*}

\begin{figure*}
    \centering
    \includegraphics[width=\textwidth]{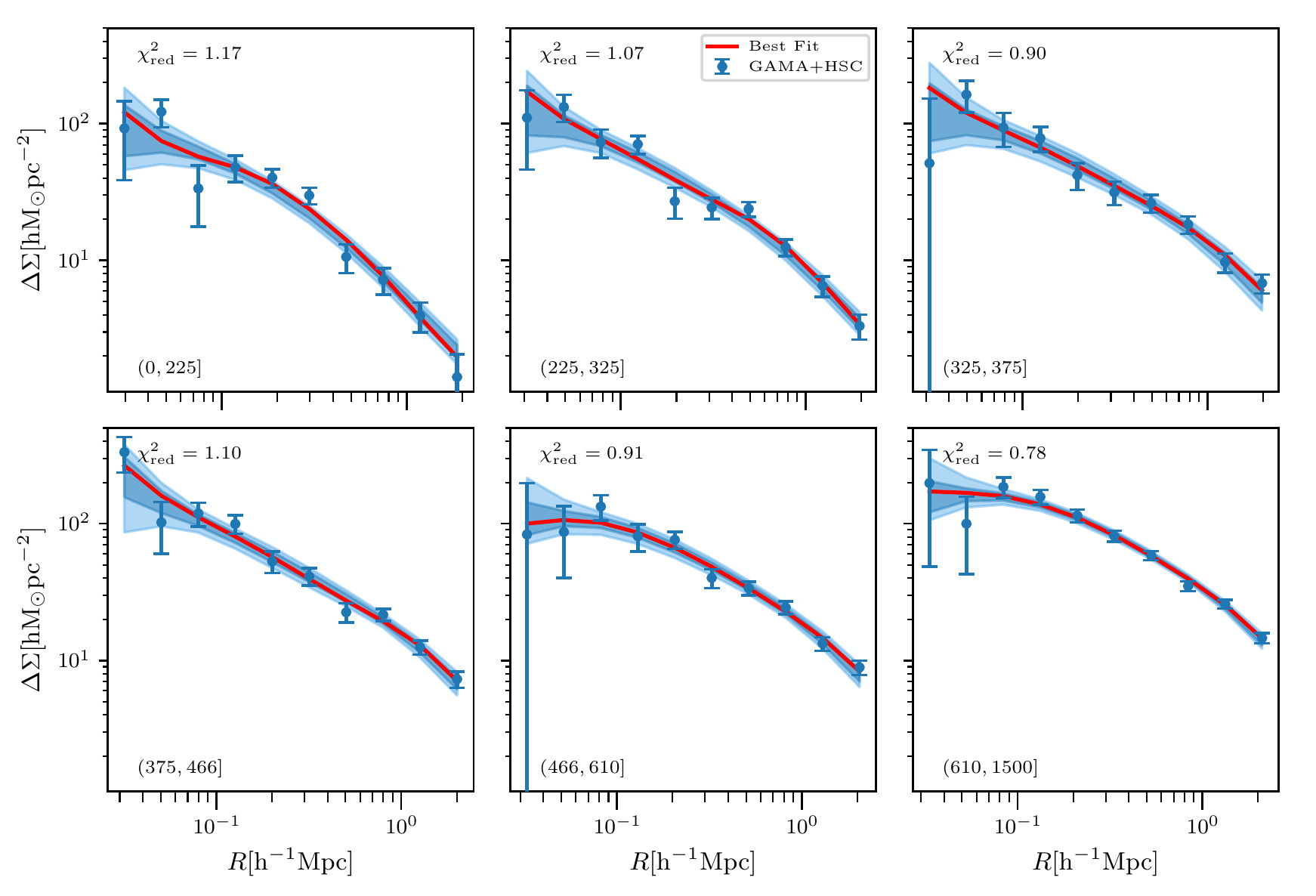}
    \caption{Same as Figure \ref{fig:modlum}, but for galaxy groups binned by group velocity dispersions. The corresponding best fit parameters can be found in Table \ref{table_4}.} 
    \label{fig:modvel}
\end{figure*}

\section{Results}
\label{sec:results}

\subsection{Lensing measurements}
\label{sec:measurements}
In this section, we will describe the results from the weak lensing analysis on GAMA galaxy groups that we carried out using the methodology discussed in the previous sections. We restrict our analysis to galaxy groups that have at least five members and those that reside within the footprint of the HSC first year shape catalogue. We divided these selected groups into six bins of galaxy group observables -- the group r-band luminosity $L_r$ and the spectroscopically determined group velocity dispersion $\sigma$ as tabulated in Table \ref{table_1}. We have measured the ESD profile in 10 comoving radial bins around the GAMA galaxy groups. Following \citet[][]{2015MNRAS.452.3529V}, we use 10 logarithmically-spaced radial bins between a comoving distance of 20 $(1+z)$ $h^{-1}$ kpc to 2 $(1+z)$ $h^{-1}$ Mpc, where $z$ denotes the mean redshift in each of the selected bins. We use the brightest cluster/group galaxy (BCG) as the group centre for the ESD measurement. We model these ESD profiles using the halo model described in section \ref{sec_3.1} and infer the mean halo masses. We then use the inferred halo masses to study their correlations with $r$-band group luminosity and velocity dispersion.

In Figures \ref{fig:modlum} and \ref{fig:modvel}, we present our modelled galaxy-galaxy lensing measurements around galaxy groups binned by their $r$-band group luminosity and velocity dispersion. The blue data points are the results from our measurements using the HSC S16a shape catalogue. We obtain a total signal to noise ratio of 55  and 51 when combined across all six different bins for two different selections. 

\begin{figure*}
    \centering
    \includegraphics[width=\columnwidth]{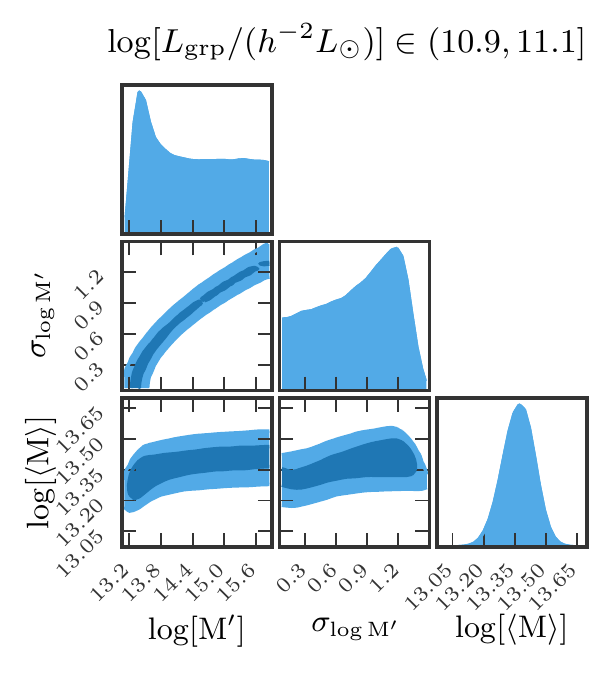}
    \includegraphics[width=\columnwidth]{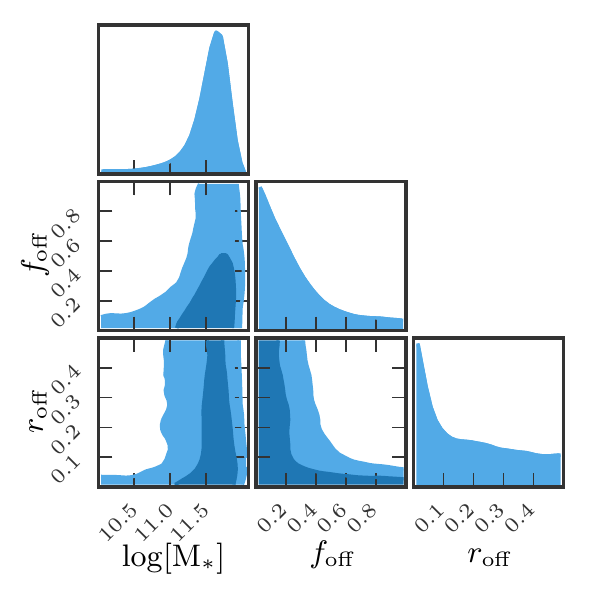}
    
    \caption{{\it Degeneracies in parameter inferences}: The left panel shows the degeneracies of the posterior distribution of halo occupation parameters and the resulting mean halo masses for galaxy groups binned in group luminosity and velocity dispersion, respectively. The right panel shows the degeneracies of the posterior distribution of our nuisance parameters corresponding to off-centering systematics and the baryonic contribution of the BCG in the galaxy group. These plots correspond to one of the bins $\log[L_{\rm grp} / (h^{-2}L_\odot)] \in (10.9,11.1]$ and is representative of the degeneracies seen in all the other bins.}
    \label{fig:post}
\end{figure*}

\begin{figure*}
    \centering
    \includegraphics[width=\columnwidth]{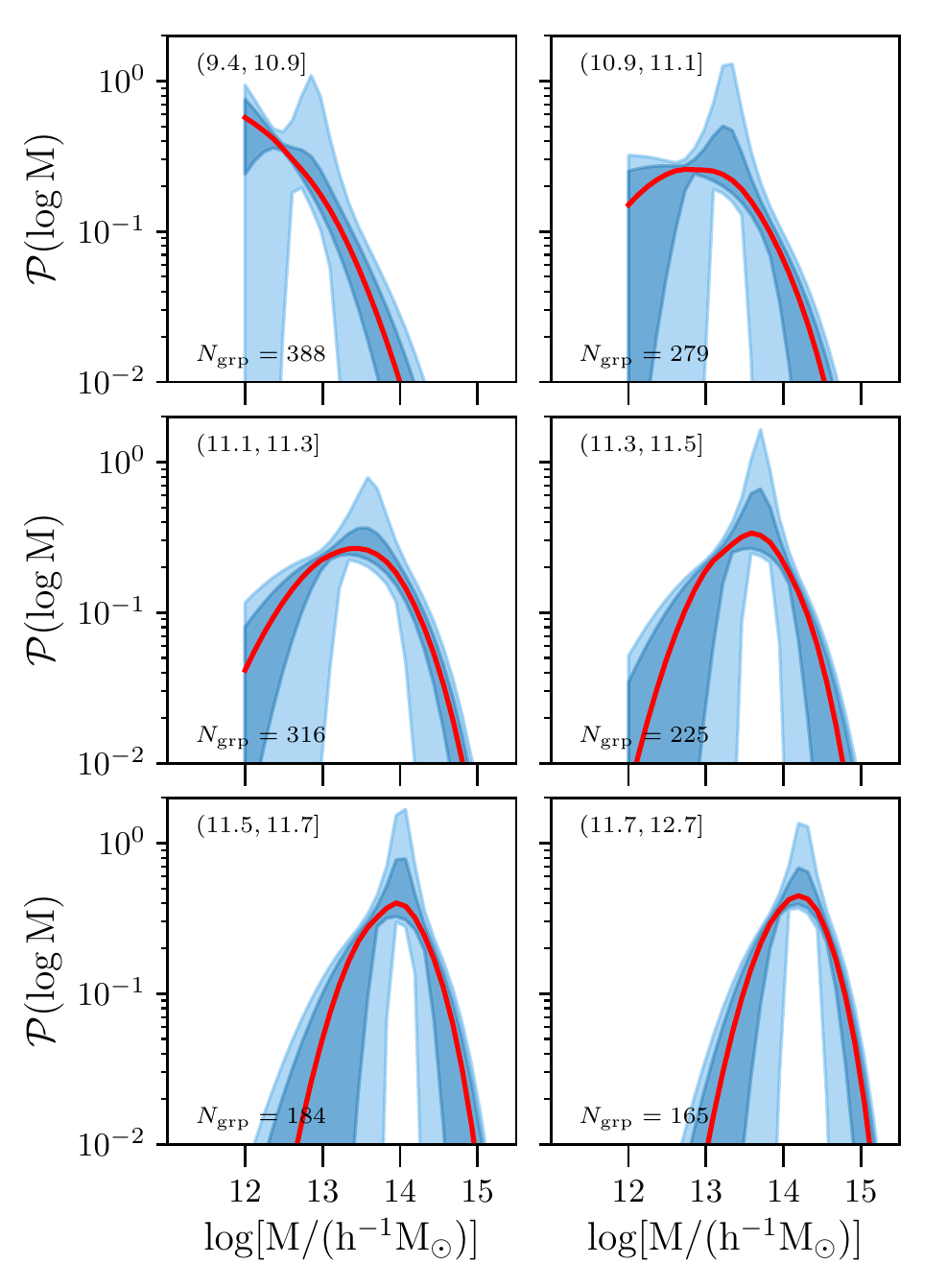}
    \includegraphics[width=\columnwidth]{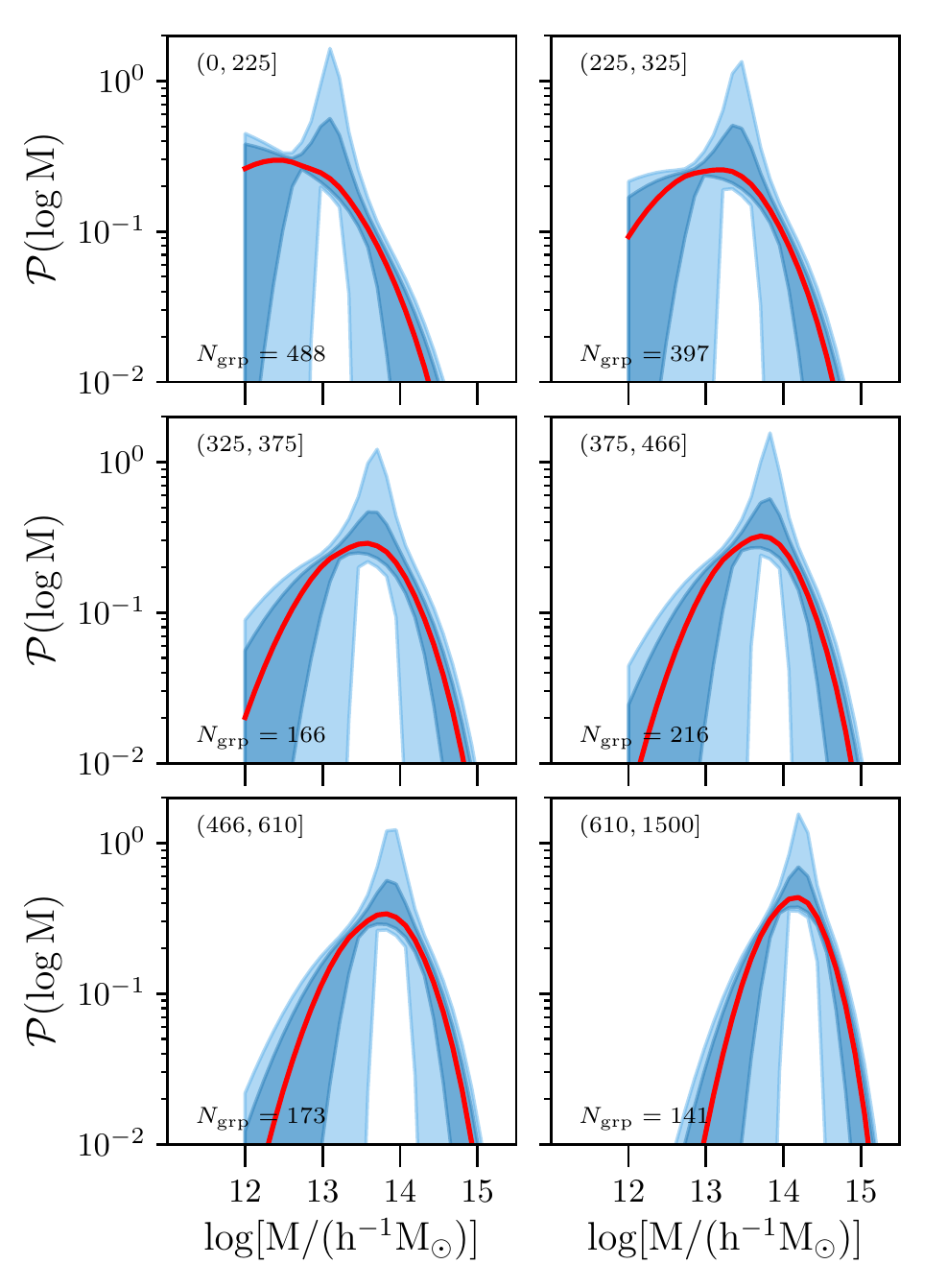}
    
    \caption{{\it Halo occupation distribution of galaxy groups:} The left and the right panel correspond to galaxy groups binned in group luminosity and velocity dispersion, respectively. The red line corresponds to the median while the dark and light shaded blue regions show the 68 and 95 percent posterior predictive distribution of halo masses given the specific bin in group observable indicated in the top left of each subpanel and number of groups $N_{\rm grp}$ for each bin in the bottom left.}
    \label{fig:plogm}
\end{figure*}

The lensing shear is related to the surface density of matter around galaxy groups. It is clear from the different panels in the figures, that the lensing shear introduced by the lens on the observed shape of the source galaxies decreases as we move further away from the group centre. Comparison of the lensing signal amplitude at about 1 $\hinvMpc$ between the different panels shows that as we go to higher group luminosity bins, we are probing more massive dark matter haloes. We see similar behaviour for the measurement of the lensing signal as a function of galaxy group velocity dispersion with more dynamically massive galaxy groups showing a stronger weak lensing signal.

\begin{table*}
\centering
\begin{tabular}{ |p{3cm}||p{2cm}|p{2cm}|p{2cm}|p{2cm}|p{2cm}|p{2cm}|}
 \multicolumn{7}{|c|}{} \\
 \hline
 Parameters &  & & $\log[L_{\rm grp}/(h^{-2} L_\odot)]\in$ & & & \\
 & (9.4,10.9] &(10.9,11.1] & (11.1,11.3] & (11.3,11.5] & (11.5,11.7] & (11.7,12.7]\\
 \hline
$\log[{\rm M'/(h^{-1} M_{\odot})}]$
&$13.22_{-0.40}^{+0.61}$
&$14.35_{-0.89}^{+1.05}$
&$14.98_{-0.83}^{+0.65}$
&$14.48_{-0.64}^{+0.92}$
&$14.76_{-0.62}^{+0.79}$
&$15.13_{-0.63}^{+0.54}$
\\
$\sigma_{\rm \log M'}$
&$1.05_{-0.54}^{+0.33}$
&$0.84_{-0.50}^{+0.36}$
&$0.81_{-0.30}^{+0.17}$
&$0.60_{-0.33}^{+0.27}$
&$0.52_{-0.32}^{+0.21}$
&$0.50_{-0.23}^{+0.13}$
\\
$f_{\rm off}$ 
&$0.22_{-0.17}^{+0.38}$
&$0.19_{-0.14}^{+0.30}$
&$0.46_{-0.12}^{+0.11}$
&$0.13_{-0.09}^{+0.19}$
&$0.27_{-0.09}^{+0.08}$
&$0.17_{-0.08}^{+0.08}$
\\
$r_{\rm off}$
&$0.12_{-0.09}^{+0.22}$
&$0.16_{-0.13}^{+0.21}$
&$0.35_{-0.13}^{+0.10}$
&$0.20_{-0.18}^{+0.20}$
&$0.32_{-0.11}^{+0.11}$
&$0.35_{-0.16}^{+0.11}$
\\
$\log[{\rm M_*/(h^{-1} M_{\odot})}]$
&$11.17_{-0.59}^{+0.29}$
&$11.61_{-0.29}^{+0.17}$
&$11.68_{-0.25}^{+0.15}$
&$11.07_{-0.69}^{+0.45}$
&$11.00_{-0.65}^{+0.53}$
&$11.35_{-0.84}^{+0.44}$
\\
\hline
$\log[{\rm \langle M\rangle/(h^{-1} M_{\odot})}]$
&$12.88_{-0.14}^{+0.11}$
&$13.37_{-0.07}^{+0.07}$
&$13.71_{-0.07}^{+0.07}$
&$13.77_{-0.06}^{+0.05}$
&$14.07_{-0.05}^{+0.05}$
&$14.29_{-0.04}^{+0.04}$
\\
SNR
&10.33
&15.26
&18.28
&23.76
&24.31
&34.39
\\

$\chi_{\rm min}^{2}/{\rm dof_{eff}}=\chi^2_{\rm red}$
&10.13/8.22
&4.27/8.25
&6.41/6.53
&7.79/7.94
&9.74/6.69
&7.21/7.01
\\
\hline
\end{tabular}
\caption{{\it Parameter constraints}: The table lists the median and the errors (based on the 16 and 84 percentiles of the posterior distribution) of the model parameters given the weak lensing measurements for galaxy groups binned by group luminosity$\log[L_{\rm grp}/(h^{-2} L_\odot)]$. The table also lists the inferred mean halo mass $\log[\avg{M}(\hinvMsun)]$, and the signal-to-noise ratio SNR for the weak lensing measurements. We also list the $\chi^2_{\rm red}$ for each of the bins.}

\label{table_3}
\end{table*}

\begin{table*}
\centering
\begin{tabular}{ |p{3cm}||p{2cm}|p{2cm}|p{2cm}|p{2cm}|p{2cm}|p{2cm}|}
 \multicolumn{7}{|c|}{} \\
 \hline
 Parameters &  & & $\sigma/(\kms)\in$ & & & \\
  & (0,225] &(225,325] & (325,375] & (375,466] & (466,610] & (610,1500]\\
 \hline
$\log[{\rm M'/(h^{-1} M_{\odot})}]$
&$14.06_{-0.82}^{+1.08}$
&$14.54_{-0.92}^{+0.95}$
&$14.80_{-0.83}^{+0.76}$
&$14.78_{-0.75}^{+0.78}$
&$14.88_{-0.74}^{+0.71}$
&$15.04_{-0.60}^{+0.57}$
\\
$\sigma_{\rm \log M'}$
&$0.89_{-0.59}^{+0.45}$
&$0.81_{-0.47}^{+0.29}$
&$0.72_{-0.35}^{+0.21}$
&$0.64_{-0.34}^{+0.21}$
&$0.63_{-0.31}^{+0.18}$
&$0.50_{-0.24}^{+0.14}$
\\
$f_{\rm off}$ 
&$0.22_{-0.17}^{+0.40}$
&$0.26_{-0.15}^{+0.17}$
&$0.32_{-0.14}^{+0.13}$
&$0.35_{-0.12}^{+0.11}$
&$0.25_{-0.11}^{+0.10}$
&$0.13_{-0.08}^{+0.11}$
\\
$r_{\rm off}$ 
&$0.09_{-0.07}^{+0.20}$
&$0.25_{-0.14}^{+0.15}$
&$0.32_{-0.14}^{+0.12}$
&$0.37_{-0.12}^{+0.09}$
&$0.34_{-0.15}^{+0.11}$
&$0.33_{-0.28}^{+0.13}$
\\
$\log[{\rm M_*/(h^{-1} M_{\odot})}]$  
&$11.11_{-0.61}^{+0.33}$
&$11.41_{-0.50}^{+0.24}$
&$11.24_{-0.74}^{+0.41}$
&$11.74_{-0.29}^{+0.16}$
&$10.81_{-0.55}^{+0.55}$
&$10.91_{-0.61}^{+0.60}$
\\
\hline
$\log[{\rm \langle M\rangle/(h^{-1} M_{\odot})}]$  
&$13.20_{-0.08}^{+0.07}$
&$13.50_{-0.06}^{+0.06}$
&$13.76_{-0.08}^{+0.08}$
&$13.88_{-0.06}^{+0.07}$
&$13.96_{-0.06}^{+0.06}$
&$14.26_{-0.04}^{+0.04}$
\\
SNR
&14.91
&18.35
&15.21
&20.54
&20.19
&32.00
\\

$\chi_{\rm min}^{2}/{\rm dof_{eff}}=\chi^2_{\rm red}$
&10.12/8.66
&8.15/7.61
&6.44/7.19
&7.21/6.57
&6.37/6.99
&5.81/7.49
\\
\hline
\end{tabular}
\caption{Same as Table \ref{table_3} but for galaxy groups selected according to group velocity dispersion.}
\label{table_4}
\end{table*}

\subsection{HOD model inferences}
\label{sec:constraints}
Given these ESD measurements, and their covariance matrix which is dominated by shape noise errors (see Section \ref{sec_3.2}), we proceed to obtain the posterior distribution of our model parameters given the data. We fit a HOD model described in Section \ref{sec_4} to the measurements in each of the group luminosity and velocity dispersion bins separately to infer the halo masses of these galaxy groups.

Figures \ref{fig:modlum} and \ref{fig:modvel} shows the ESD measurements along with the 68 and 95 percent credible model predictions (dark and light shaded regions) computed from the posterior distribution of the parameters given the ESD measurements for the galaxy group $r$-band luminosity and velocity dispersion selection, respectively. The red line in each of the panels corresponds to the best fit model predictions. We also indicate the $\chi^2_{\rm red}$ value for the best fit, by using the effective degrees of freedom \citep[see eqn 29 in][]{2019PhRvD..99d3506R}.

We observe that our halo model provides reasonable fits for the weak lensing signal around the groups with reasonable $\chi^2_{\rm red}$ given the effective degrees of freedom for the individual ESD measurements. Our model parameters give us the halo occupation probability $\mathcal{P}(X_{\rm grp}|M)$. We use Eqn(\ref{eq:avgmass}), to compute the posterior predictive distributions of the mean halo mass $\langle M \rangle$ for each selection bin.

In Tables \ref{table_3} and \ref{table_4}, we present the posterior distribution of our model parameters given the ESD measurements around galaxy groups binned by $L_{\rm grp}$ and $\sigma$, respectively. We summarize the posterior distribution by tabulating the median, and the errors which correspond to the 16th and 84th percentiles of the posterior distribution. In each of the tables, we also show the corresponding constraints on the mean halo mass for each selection bin. We also tabulate the signal-to-noise SNR for our ESD measurements and the best fit reduced chisq $\chi^2_{\rm red}$ values for each of the fits. We explicitly show the variation of the HOD parameters for both the galaxy group $r$-band luminosity and velocity dispersion bins. 

Given our adopted functional form for the halo occupation distribution we have allowed the scatter in halo masses to vary for each of our fits. Although the errors are large we see a systematic trend of decreasing $\sigma_{\rm log M^\prime}$ as a function of group luminosity and velocity dispersion. These values can be compared to the posterior distribution of $\sigma_{\rm log M^\prime} = 0.74_{-0.16}^{+0.10}$ obtained by \citet[][]{2015MNRAS.452.3529V} who assumed a fixed value for $\sigma_{\rm \log M^\prime}$ in their analysis of GAMA galaxy groups binned by total $r$-band group luminosity.

As an example, in Figure \ref{fig:post}, we show the posterior distribution of the model parameters for one of our $r$-band group luminosity selection bins ($\log[L_{\rm grp} / (h^{-2} L_\odot)] \in (10.9,11.1]$) with 68 and 95 percent credible intervals. In the left panel, we present the degeneracies between the model parameters related to the halo occupation distribution and the derived parameter ${\rm \log \langle M \rangle}$. Given the functional form of our HOD, there is an expected degeneracy between the parameters ${\rm \log M'}$ and $\sigma_{\rm \log M'}$, where a large value of one parameter can be traded off by increasing the other parameter. However, the sub-panels in the bottom row show that the mean halo mass $\log \langle M \rangle$ is quite robustly measured despite these degeneracies. 

In the right panel, we show the degeneracies in the nuisance parameters of our model, the stellar mass component ${\rm M_*}$, and the off-centering parameter $f_{\rm off}$ and $r_{\rm off}$. Once again we observe expected degeneracies between the off-centering parameters, where larger values of the off-centering fractions $f_{\rm off}$ can be tolerated only if the off centring kernel is not too broad. On the other hand larger values of the off-centring kernel can only correspond to a smaller fraction of such off-centred groups. We do not observe any strong degeneracies between the other model parameters and $\log \langle M \rangle$. Even though we only show these parameter degeneracy plots for only one group luminosity bin, we observe similar features in other bins.

In Figure \ref{fig:plogm}, we present the posterior predictive distribution of $P(\log M|X_{\rm grp})$ for galaxy groups (see Eq.~\ref{eq:avgmass}) binned by their group luminosities and group velocity dispersions, respectively. Most of the distributions show a well defined peak corresponding to the average mass of the haloes the galaxy groups occupy. These probability distributions move to higher and higher mass haloes as we consider larger group luminosity and velocity dispersion selections. In the faintest group luminosity bin and the smallest velocity dispersion bin, we see that the distribution is cut off at the low halo mass end at a mass scale of $10^{12} h^{-1} M_\odot$. This is a consequence of the mass resolution to which we trust our modelling with the \textsc{DARK EMULATOR}. We explore the consequence of this mass resolution threshold in \textsc{DARK EMULATOR}, by comparing the results to those obtained using the analytical HOD modelling code \textsc{AUM} \citep{2013MNRAS.430..725V, 2013MNRAS.430..747M, 2013MNRAS.430..767C}, which does not have such a resolution limit but relies on simplified analytical descriptions of the halo matter cross-correlation. In Appendix~\ref{app:aum}, we show that the average halo masses we obtain from the two modelling schemes are quite consistent with each other.

\subsection{Galaxy group observable scaling relations}
\label{sec:scaling}

Given the mean halo masses as a function of the group luminosity and the group velocity dispersion, we now proceed to obtain the scaling relation between galaxy group observable and their halo masses. The scaling relations will prove to be the first stepping stone to study galaxy formation and evolution in galaxy groups.

We explore simple power law models between the galaxy group observable and halo mass similar to the scaling relation parameterization adopted by \citet[see Eqn 37 in][]{2015MNRAS.452.3529V},
\begin{equation}
\frac{\avg{M}}{10^{14}\hinvMsun} = A \left(\frac{X_{\rm grp}}{X_{\rm piv}}\right)^\alpha\,,\label{scaling}   
\end{equation}
where $A$ is the amplitude of the scaling relation, and $\alpha$ is the power law index of the scaling relation, and $X_{\rm piv}$ is the pivot group observable we choose for our scaling relation. We carry out a Bayesian inference of the parameters of the scaling relation given the measured mean halo masses for group observables in a given bin. We fit the power-law scaling relations as linear regression in log basis with a flat prior of [0,10] for both $A$ and $\alpha$ \citep[][]{2015MNRAS.452.3529V}. The likelihood of these measurements given the model parameters (amplitude and exponent) is given by
\begin{equation}
    \mathcal{L} = \prod_i \exp\left( -  \frac{\left[ \log \avg{M}_i - \log \avg{M}_{\rm scale, i} \right]^2 }{2 \sigma^2_{i}} \right)\,,
\end{equation}
where $\log[\avg{M}]_i$ denotes the logarithm of the average masses inferred from weak lensing for the $i$-th bin in a group observable property, and $\sigma_{i}$ denotes the error on the inferred values of $\log[\avg{M}]$. The quantity $\log\avg{M}_{\rm scale, i}$ denotes the value predicted by the scaling relation parameters. In order to compute this quantity, we need to account for the distribution of the group observables in a given bin in order to avoid biases due to intrinsic distribution of the group observable within a bin \citep{Kelly2007}. Therefore, we assign the halo mass to each group in a given bin according to the scaling relation and compute the average mass in each selection bin.

Given that each of the inferences of the mean halo masses are obtained independently of each other, we take the total likelihood to be a product over the six group observable bins. We sample the posterior distribution of the scaling relation parameters given our measurements of the mean halo masses using a procedure similar to the one detailed in Section~\ref{sec:theory}.

The scaling relations obtained in the above manner allow a direct comparison with \citet[][]{2015MNRAS.452.3529V}. However, we also show the results for the scaling relation
\begin{equation}
\avg{\log M_{14}} = \log \tilde{A}  + \tilde \alpha \log \left(\frac{X_{\rm grp}}{X_{\rm piv}}\right)\,,\label{scaling:log} 
\end{equation}
where $M_{14}=M/10^{14}\hinvMsun$. This scaling relation corresponds to $\avg{\log M}$ in contrast to $\avg{M}$ in Eq.~\ref{scaling}. We also attempt to constrain the intrinsic scatter in this scaling relation, $\sigma^{\rm int}_{\log M}(X_{\rm grp})$. We assign halo masses to each of the groups in a given bin according to the above scaling relation and include a log-normal scatter around it. We compute $\avg{M}$ and $\sigma_{\log M}$ within each bin accounting for the distribution of group observables in each bin as well as the intrinsic scatter. Since both these quantities are inferred from our weak lensing analyses, we use these constraints to infer the posterior distribution of $\tilde A$, $\tilde \alpha$ as well as $\sigma^{\rm int}_{\log M}$. We include flat uninformative priors between $[0, 10]$ for $\tilde{A}$, $\tilde\alpha$, and $[0,2]$ for $\sigma^{\rm int}_{\log M}$.

\begin{figure}
    \centering
    \includegraphics[width=\columnwidth]{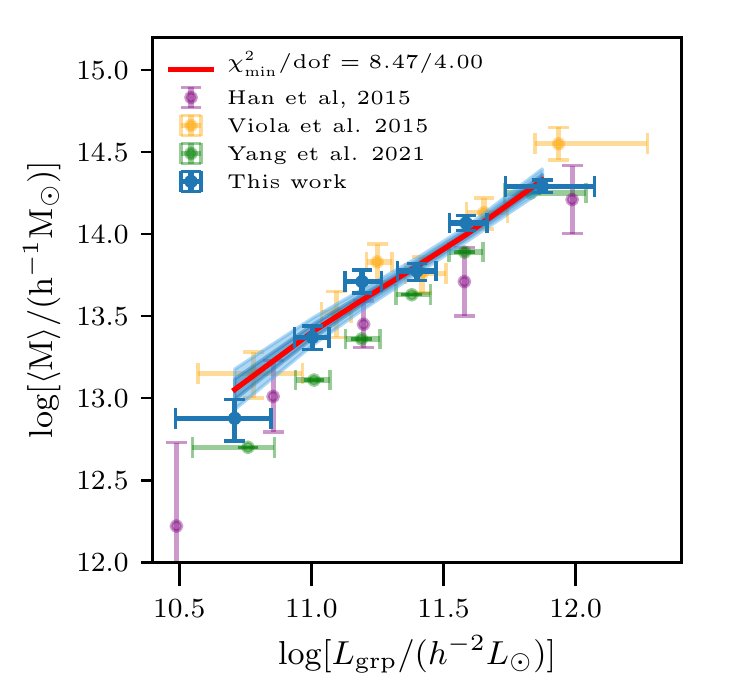}
    \caption{{\it The halo mass-group luminosity scaling relation:} The blue points with errors represent the inferred mean halo masses from our analysis of the weak lensing measurements of galaxy groups. The red line shows our best fit power-law model (labeled with its $\chi^2_{\rm red}$ in the legend), and the dark and light blue shaded regions correspond to the 68 and 95 percent predictions based on our power-law model fit to our measurements. For comparison, we show the previous results from \citet[][]{2015MNRAS.452.3529V} using the KiDS shape catalogue with orange points. The purple data points denotes the scaling relation measurements from fig.~3 of \citet[][]{2015MNRAS.446.1356H}, who used SDSS source galaxies around GAMA groups with at least three members.  The green points are the estimates of the mean halo masses based on abundance matching of galaxy groups from the DESI legacy imaging survey \citep[][]{2021ApJ...909..143Y}. } 
    \label{fig:masslum}
\end{figure}
\begin{figure}
    \centering
    \includegraphics[width=\columnwidth]{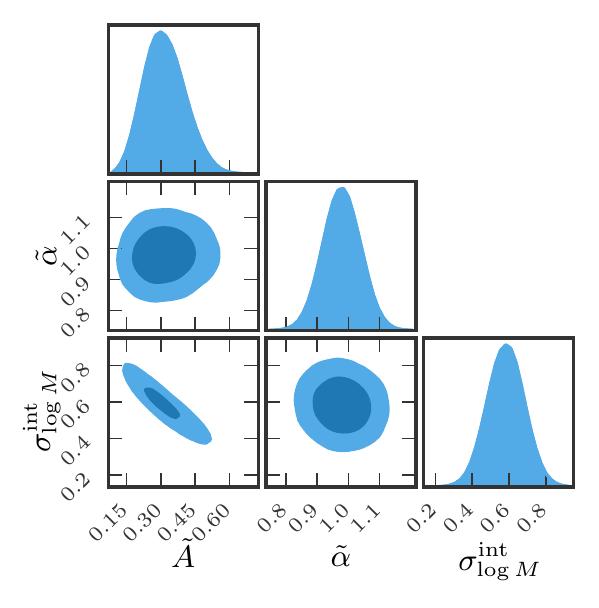}
    \caption{{\it Posterior distribution for the halo mass-group luminosity with an intrinsic scatter:} We show the posterior distributions for the amplitude $\tilde{A}$, slope $\tilde{\alpha}$ as described by Eq. \ref{scaling:log} for group luminosity-halo mass scaling relation and the intrinsic scatter $\sigma^{\rm int}_{\log M}$ for this relation. }
    \label{int_lum}
\end{figure}
\subsubsection{Group Luminosity and Halo Mass Relation}
\label{lum_mass}
In Figure \ref{fig:masslum}, the blue points with errors are the results from our HSC weak lensing analysis of the GAMA galaxy groups binned in group $r$-band luminosity and correspond to the median values and their $1\sigma$ errors. The red line corresponds to the best fit scaling relation with a $\chi^2_{\rm red} = 2$ for approximately 4 effective degrees of freedom. The dark and light blue bands correspond to 68 and 95 percent credible intervals around the median model predictions.

Our results correspond to a power-law scaling relation between the halo mass at fixed group luminosity given by
\begin{equation}
\frac{\avg{M}}{10^{14}{\,\rm h^{-1} M_\odot}} = (0.81\pm 0.04)\left(\frac{L_{\rm grp}}{10^{11.5}{\, \rm h^{-2} L_{\odot}}}\right)^{(1.01\pm0.07)}\,.    
\end{equation}
Our simple power law model can explain the inferred values of halo masses and gives some of the tightest constraints on the scaling relation of halo mass and group luminosity. The scaling relation between halo mass and BCG luminosity typically has a much steeper slope and a larger scatter at fixed BCG luminosity \citep[see e.g.,][]{2011MNRAS.410..210M}. 
For comparison, we also show the corresponding results from \citet{2015MNRAS.452.3529V} as orange data points with errors. Qualitatively, both the measurements follow a similar trend as the mean halo mass increases with group $r$-band luminosity while spanning two orders of magnitude. It implies that groups of higher luminosities tend to live inside haloes of larger masses. Quantitatively, the power law index of the scaling relation that we obtain ($1.01 \pm 0.07$) agrees with the power law index ($1.16 \pm 0.13$) obtained by \citet[][]{2015MNRAS.452.3529V} but with significantly smaller errors. The value of the amplitude of the scaling relation that we obtain $A=0.81\pm 0.04$ is smaller yet consistent with the amplitude obtained in that study, $A=0.95 \pm 0.14$, but with significantly smaller errors. These smaller errors are primarily due to the larger galaxy number density and overlapping area in our analysis than previous results.

In Figure \ref{fig:masslum}, we also compare our results to those obtained by \citep{2015MNRAS.446.1356H} who studied GAMA galaxy groups using maximum likelihood weak lensing method with source galaxies from the Sloan Digital Sky Survey (SDSS). In their study, the authors use GAMA groups with at least 3 members (compared to 5 members used in this work). The amplitude of the scaling relation obtained in their study is smaller than what we obtain by about $0.16$ dex, although the error on the amplitude quoted in their study is about $0.12$ dex. This is likely a result of the further dependence of the scaling relation on the group multiplicity/richness. The slope of the scaling relation they obtain is consistent with our results.

Finally we also compare our results to the abundance matching technique applied on galaxy groups constructed from the DESI legacy imaging surveys galaxy group catalogue \citep[][]{2021ApJ...909..143Y}. We used the group richness, the group halo mass, the group z-band luminosities and the redshift information provided in the catalogue, and applied similar selection cuts within the redshift range $z < 0.5$ as we have applied to our lensing sample. The green data points show the mean values with $1\sigma$ errors estimated from this DESI DR8 group catalogue. Given the differences in survey depth and input cosmology, a proper quantitative comparison especially with the amplitude of the scaling relation cannot be made. Qualitatively, we do see a similar behavior to that of our scaling relation for groups with $L_{\rm grp} > 10^{11.3} h^{-2}L_\odot$, with some hints of a steeper relation at lower group luminosities.

We have also obtained constraints on the scaling relation $\avg{\log M_{14}}$ given in Eq.~\ref{scaling:log} and its intrinsic scatter according to the procedure described in Section~\ref{sec:scaling}. We obtain the posterior distributions of $\tilde{A}$, $\tilde{\alpha}$ and $\sigma_{\log M}^{\rm int}$ given our weak lensing inferences. These posterior distributions are shown in Fig.~\ref{int_lum}. We notice very little correlation between the inference of the amplitude and the slope of the scaling relation. We also see the expected anti-correlation between the amplitude and the intrinsic scatter. In general, for a log-normal distribution, $\avg{\log M}$ is smaller than $\log\avg{M}$ with a difference that approaches zero as the intrinsic scatter approaches zero. Our weak lensing constraints yield $\avg{M}$. This implies that a larger scatter will require a smaller amplitude for the scaling relation between $\avg{\log M}$ and group luminosity in order to match the weak lensing constraints.

We obtain $\tilde A=0.30\pm0.09$, $\tilde \alpha=1.1\pm0.1$, and an intrinsic scatter of $\sigma_{\log M}^{\rm int}=0.6\pm{0.1}$. Although at face value, this difference in the amplitude between the two analyses seems large, this is entirely due to the fact that the two analyses characterize the scaling relation using separate statistical measures, $\log\avg{M}$ versus $\avg{\log M}$, and these characterizations differ from each other in the presence of a scatter in the scaling relation. The large value of scatter we obtain is consistent with $\sigma_{\log M}^{\rm int}\sim 0.5$ reported based on the analysis of GAMA mock catalogues (albeit based on a previous version of the group finding algorithm) by \citet[][]{2015MNRAS.446.1356H}.

The difference between the true halo mass and the weak lensing inferred mass due to projection effects can cause $20$ percent scatter \citep[see e.g.,][]{BeckerKravtsov2011}, and thus is not the dominant source of the scatter that we observe in the scaling relation. The large value of the scatter we observe is likely a result of the intrinsic scatter in the scaling relation between halo mass and the galaxy group observable, as well as the scatter introduced by the identification algorithm for galaxy groups. The algorithm could cause some actual galaxy groups to be fractured into parts or some groups to be joined together even though they are separate \citep[see e.g.][in the context of SDSS]{Campbell2015}, thus causing an increase in the scatter in the scaling relation. Accounting for such effects will require detailed study using mock catalogs, which we defer future, once such mock catalogs become available. The presence of a large scatter makes the mass determination of individual galaxy groups based on their observable group luminosity subject to large uncertainties. We further note that \citet{2015MNRAS.446.1356H} had to assume a value of this scatter based on inputs from mock catalogs in order to infer the scaling relation of galaxy groups, whereas we are able to constrain both the scaling relation and its scatter directly from the data.

\begin{figure}
    \centering
    \includegraphics[width=\columnwidth]{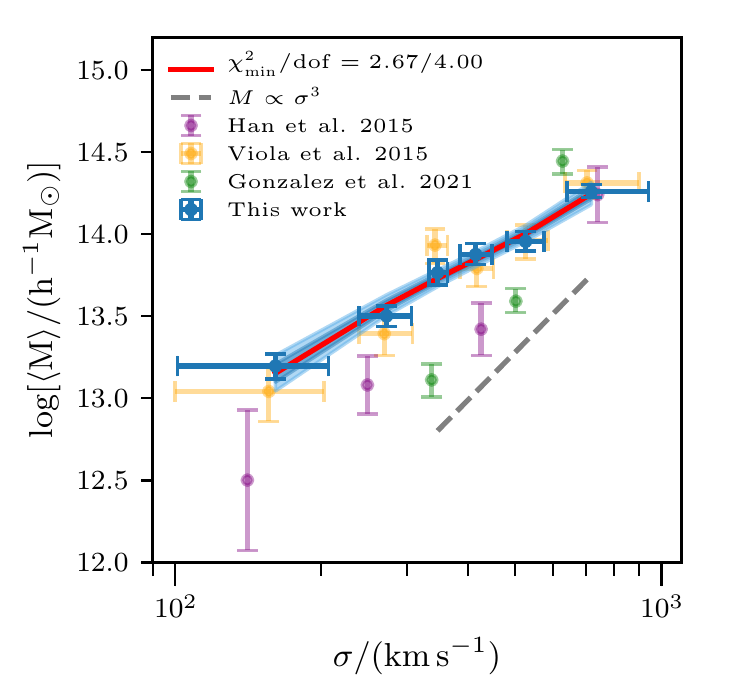}
    \caption{{\it The halo mass-group velocity dispersion scaling relation:} The blue points with errors represent the inferred mean halo masses from our analysis of the weak lensing measurements of galaxy groups. The red line shows the best fit power-law model (labeled as $\chi^2_{\rm red}$ in the legend) with dark and light blue shaded regions that correspond to the 68 and 95 percent predictions based on our power-law model fit to our measurements. For comparison, we show the results from \citet[][]{2015MNRAS.452.3529V} using the KiDS weak lensing measurements as orange points with errors. The purple data points denote the scaling relation measurements from fig.~4 of \citet[][]{2015MNRAS.446.1356H} who used SDSS source galaxies around GAMA groups with at least three members. The green points correspond to more recent weak lensing mass estimates from \citep[][]{2021MNRAS.504.4093G} for galaxy group sample constructed from the galaxies in SDSS-DR12 and having at least four member galaxies. }
    \label{fig:massvel}
\end{figure}
\begin{figure}
    \centering
    \includegraphics[width=\columnwidth]{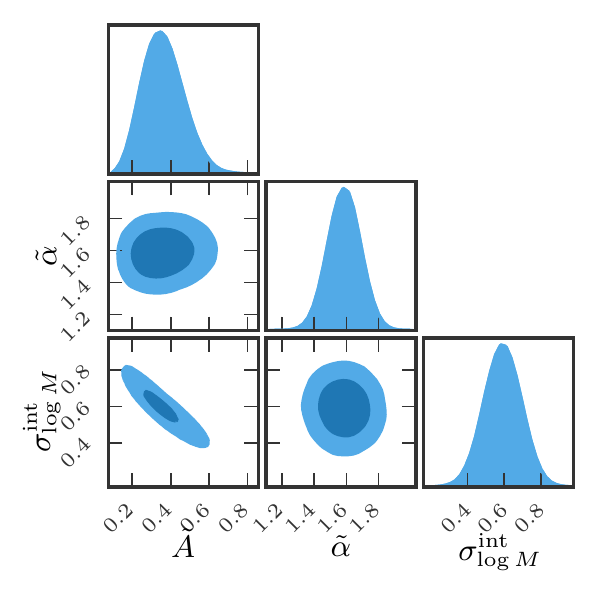}
    \caption{{\it Posterior distribution for the halo mass-velocity dispersion scaling relation with an intrinsic scatter:} We show the posterior distributions for the amplitude $\tilde{A}$, slope $\tilde{\alpha}$ as described by Eq. \ref{scaling:log} for halo mass-group velocity dispersion scaling relation and the intrinsic scatter $\sigma^{\rm int}_{\log M}$ for this relation.}
    \label{int_vel}
\end{figure}
\subsubsection{Velocity Dispersion and Halo Mass Relation}
We follow a similar methodology for computing the halo mass velocity dispersion scaling relation of galaxy groups as we used in the last section. We have divided our lens samples into six bins of velocity dispersions (see Table \ref{table_1}). The ESD signals that we measure from our data are shown for each of the velocity dispersion bins as blue points with errorbars in Figure \ref{fig:modvel}. We modelled these ESD measurements with the HOD framework described in Section \ref{sec_4}. The HOD model is a good description of the lensing data. The posterior distribution of the HOD parameters given the data can be used to compute the mean halo masses for each of our bins.

These inferred masses as a function of the velocity dispersion are shown as blue data points with errors in Figure~\ref{fig:massvel}. We model this scaling relation as a simple power-law following the parameterization adopted in \citet[][]{2015MNRAS.452.3529V}. The solid red line denotes the best fit model which has a $\chi^2_{\rm red} =0.66$. The dark and light blue shaded regions show the 68 and 95 percent intervals of the predictions for the scaling relation given our model and the posterior distributions of the scaling relation parameters. 

The scaling relation between halo mass and velocity dispersion obtained from our analysis can be summarized as
\begin{equation}
\frac{\avg{M}}{10^{14}{\, \rm h^{-1} M_\odot}} = (0.93 \pm 0.05)\left(\frac{\sigma}{500\,\kms}\right)^{(1.52\pm0.10)}\,. \end{equation}
with a 6 percent measurement of both the amplitude and slope of the scaling relation. The slope of the scaling relation is significantly shallower than that expected from the condition of virial equilibrium which would suggest $M\propto \sigma^{3}$, as found in dissipationless numerical simulations \citep[][]{2008ApJ...672..122E, 2013ApJ...779..159D}. Previous observational studies also quote similar shallow scaling relation for the halo mass-velocity disperision for group scale objects with a power law index of $\sim 2$ \citep[see e.g.,][]{2015MNRAS.446.1356H,2015MNRAS.452.3529V}.

For comparison, we present the results of \citet[][]{2015MNRAS.452.3529V} as orange points with errors for the same velocity dispersion bins. Similar to the halo mass group luminosity scaling relation, we obtain a consistent but significantly well measured scaling relation. In their study, \citet{2015MNRAS.452.3529V} argue that the shallow slope of the scaling relation could be a result of the apparent richness cut of five or more member galaxies that we have used for the selection of GAMA groups to perform such an analysis. This would imply that at fixed velocity dispersion, groups with higher richness occupy more massive haloes. Using a dark matter only GAMA mock simulation \citep[][]{2011MNRAS.416.2640R, 2013MNRAS.429..556M}, they showed that such a selection can result in a shallower relation. However, they also note that there could be some effect from dynamical processes that can influence the scaling relations as seen in the hydrodynamical simulations \citep[][]{2013MNRAS.430.2638M}. For comparison in the same figure, we also show results from the study of \citet[][]{2015MNRAS.446.1356H} who use groups with at least 3 members. Similar to the group luminosity halo mass scaling relation, here to we find a difference in amplitude between their study and ours which is likely a result of this cut in group membership. We defer detailed investigations of such effects using mocks and hydrodynamical simulations for future work.

We also compare our results with a recent study of \citet[][]{2021MNRAS.504.4093G} who identify groups using spectroscopic galaxies from the SDSS-DR12 survey \citep[][]{2020A&A...636A..61R} and carry out a weak lensing study using a heterogenous collection of shear catalogues - CFHTLenS, CS82, RCSLens and KiDS. The green data points represent their median halo-mass constraints with $1\sigma$ error bars at a median velocity dispersion value in the corresponding bin. These measurements seem to favour a steeper relation compared to our results and other results in the literature. The authors have argued that their group catalogue and the velocity dispersion measurements could be affected by the presence of interlopers, especially at their low velocity dispersion end, which could bias the mass estimates to be lower at fixed velocity dispersion. A more quantitative comparison with our results will require a proper accounting of the differences in group definition, halo mass definitions and the modelling methodology used in their work.

Our analysis shows that the group velocity dispersion has a tight correlation with the underlying mass of the halo. The scaling relation could suffer from biases due to the selection on the basis of richness, which requires further analysis on mock catalogues. We defer the analysis of the weak lensing of galaxy groups at fixed velocity dispersion but in richness bins to a future study. With these caveats in mind, the analysis presented in this paper gives the tightest observational constraints on the halo mass-velocity dispersion relation. 

Similar to the group luminosity case, we further constrain the scaling relation between $\avg{\log M}$ and $\sigma$ using the average masses and the dispersion $\sigma_{\log M}$ obtained from our weak lensing analysis. The posterior distribution of the amplitude, the slope and the intrinsic scatter in this relation is shown in Fig.~\ref{int_vel}. We obtain $\tilde A=0.4\pm0.1$, $\tilde \alpha=1.62\pm0.15$, and an intrinsic scatter of $\sigma_{\log M}^{\rm int}=0.6\pm{0.1}$. The difference in the amplitude $A$ and $\tilde A$ as obtained in this section is also expected given that the former corresponds to the amplitude of $\avg{M}$, while the latter corresponds to $\avg{\log M}$. The large value of scatter we obtain is very much consistent with the values obtained from GAMA mock catalogs of $\sigma_{\log M}^{\rm int}\sim 0.7$ as measured by \citet{2015MNRAS.446.1356H}. 

\section{Summary}
\label{sec:conclusions}
We have used weak gravitational lensing of background galaxies by galaxy groups to investigate the connection between group observable properties such as the total group luminosity and the velocity dispersion of group member galaxies and the underlying mass of the dark matter halo. For this purpose we made use of a gravitational lens sample that consists of spectroscopically identified galaxy groups from the GAMA galaxy survey. We used source galaxies from the first year shape catalogue from the Subaru HSC survey in order to perform our analysis. A summary of our results is as follows.
\begin{itemize}
    \item We measured the weak lensing signal of GAMA galaxy groups with at least 5 members, in 6 bins each of the galaxy group luminosity and the group velocity dispersion, respectively. We obtained a total signal-to-noise ratio of $\sim 55$ and $\sim 51$ for our measurements for galaxy groups binned by group luminosity and group velocity dispersion, respectively.
    \item We interpret these measurements in the framework of the halo model which describes the halo occupation distribution of these galaxy groups. Our theoretical model relies on the use of the \textsc{Dark Emulator} for predictions of the statistical properties of the matter distribution. Our model is able to explain the observables given the current precision of the measurements.
    \item We used the posterior distribution of the halo occupation parameters to obtain the scaling relation between the mean halo mass of galaxy groups with group luminosity and group velocity dispersion. We showed that the inferred measurements of masses are robust to systematic modelling effects of mis-centering. 
    \item Our inferred scaling relations are consistent with a power law description between halo mass and group observables - group luminosity and group velocity dispersion, broadly consistent with previous studies of these scaling relations in the literature, however with significantly smaller errors.
    \item We obtain a $\sim5$ percent constraint on the amplitude of the mass group luminosity scaling relations with $\avg{M} = (0.81\pm0.04) \times 10^{14} \hinvMsun$ at a group luminosity of $L_{\rm grp}=10^{11.5}\hinvsqLsun$, and a $\sim 7$ percent constraint on its power law index $\alpha=1.01\pm0.07$.
    \item We also obtain a $\sim 5$ percent constraint on the amplitude of the mass group velocity dispersion scaling relation with $\avg{M} = (0.93\pm0.05) \times 10^{14} \hinvMsun$ at a group velocity dispersion of $\sigma=500 \kms$, and a $\sim$7 percent constraint on its power law index $\alpha=1.52\pm0.10$.
    \item We have further obtained the parameters of the scaling relations $\avg{\log M}-\log L_{\rm grp}$ and $\avg{\log M}-\log \sigma$ as well as their intrinsic scatters. A model with a constant intrinsic scatter of $0.6\pm0.1$ describes both the scaling relations given the current errors.
\end{itemize}

Our study provides a significant improvement in the weak lensing mass measurements of galaxy groups binned by group luminosity and velocity dispersion. We note that the scaling relation we derived above may be different from the true intrinsic scaling relations of galaxy groups without any richness cuts and the true galaxy group properties (rather than the observed ones which are noisy measurements of the latter) \citep[see e.g.,][]{Sereno:2015, Mantz:2016}. This requires the use of mock galaxy catalogs such as those presented in \citep{2011MNRAS.416.2640R} for the GAMA galaxy group catalog used in this paper. One could imprint a particular intrinsic scaling relation in the mock, and constrain its parameters by using a full forward model to fit $\avg{M}$ as a function of group properties as presented here.

As the signal-to-noise ratio of weak lensing measurements continue to improve further in the future, it will enable studies of the multivariate scaling relations between the group luminosity of galaxy groups, their velocity dispersions as well as the richness of these groups and thus enable a more complete understanding of the scaling relation between galaxy groups and their host halo masses.


\section*{Acknowledgements}
We thank Navin Chaurasiya, Daniel Farrow, Jiaxin Han, Amit Kumar, Konrad Kuijken, Jochen Liske, Jonathan Loveday, Preetish K.~Mishra and Peder Norberg for comments on a draft version of the manuscript and useful discussions regarding this work. We thank the anonymous referee for useful comments which helped to improve the analysis presented in the paper. DR would like to thank the University Grants Commission (UGC), India, for financial support as a senior research fellow. This work was also supported in part by MEXT/JSPS KAKENHI Grant Numbers JP19H00677, JP20H05861, JP20H01932 and JP21H01081 (TN). We also acknowledge financial support from Japan Science and Technology Agency (JST) AIP Acceleration Research Grant Number JP20317829 (TN, HM).

The corner plots in this paper are prepared using the open python package - pygtc \citep[][]{2016JOSS....1...46B}. AUM is a publically available package at https://github.com/surhudm/aum. We acknowledge the use of Pegasus, the high performance computing facility at IUCAA. 

The Hyper Suprime-Cam (HSC) collaboration includes the astronomical communities of Japan and Taiwan, and Princeton University.  The HSC instrumentation and software were developed by the National Astronomical Observatory of Japan (NAOJ), the Kavli Institute for the Physics and Mathematics of the Universe (Kavli IPMU), the University of Tokyo, the High Energy Accelerator Research Organization (KEK), the Academia Sinica Institute for Astronomy and Astrophysics in Taiwan (ASIAA), and Princeton University.  Funding was contributed by the FIRST program from the Japanese Cabinet Office, the Ministry of Education, Culture, Sports, Science and Technology (MEXT), the Japan Society for the Promotion of Science (JSPS), Japan Science and Technology Agency  (JST), the Toray Science  Foundation, NAOJ, Kavli IPMU, KEK, ASIAA, and Princeton University.

This paper is based [in part] on data collected at the Subaru Telescope and retrieved from the HSC data archive system, which is operated by Subaru Telescope and Astronomy Data Center (ADC) at NAOJ. Data analysis was in part carried out with the cooperation of Center for Computational Astrophysics (CfCA), NAOJ.

GAMA is a joint European-Australasian project based around a spectroscopic campaign using the Anglo-Australian Telescope. The GAMA input catalogue is based on data taken from the Sloan Digital Sky Survey and the UKIRT Infrared Deep Sky Survey. Complementary imaging of the GAMA regions is being obtained by a number of independent survey programmes including GALEX MIS, VST KiDS, VISTA VIKING, WISE, Herschel-ATLAS, GMRT and ASKAP providing UV to radio coverage. GAMA is funded by the STFC (UK), the ARC (Australia), the AAO, and the participating institutions. The GAMA website is http://www.gama-survey.org/ .

\section*{Data availability}

The measurements of the HSC weak lensing signal around GAMA galaxy groups binned by group luminosity and velocity dispersion are available at https://github.com/menobista/gama-hscs16a.




\bibliographystyle{mnras}
\bibliography{ms} 




\appendix
\section{Boost Parameters}
\label{app:boost}
For the measurements of the weak lensing signal of galaxy groups, we make use of the photometric redshifts of source galaxies. The deep photometry from HSC reduces the errors on the photometric redshifts but systematic calibration issues could still plague the measurements. In particular if we inadvertently use galaxies which overlap in redshift or are in the foreground compared to the lens sample, the weak lensing signal can get diluted. 

\begin{figure*}
    \centering
    \includegraphics[width=\columnwidth]{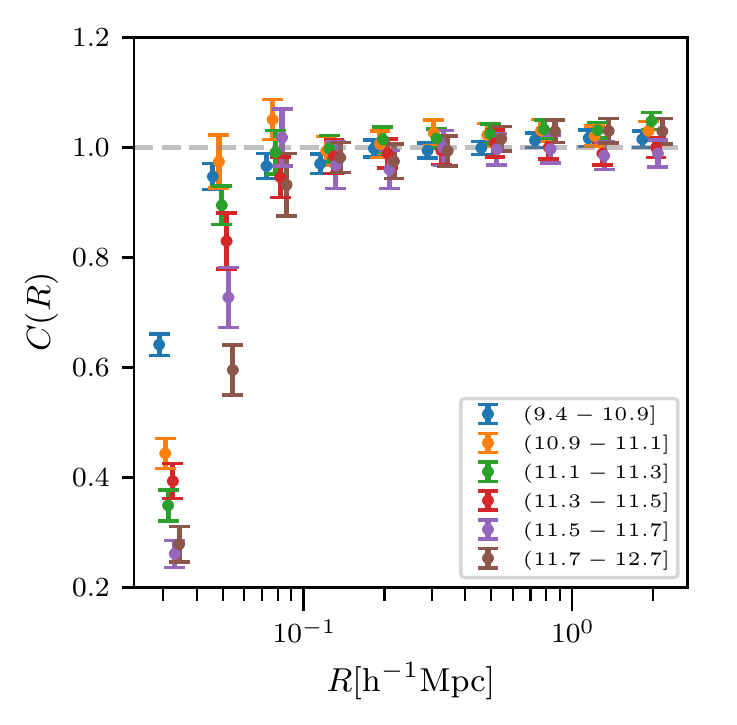}
    \includegraphics[width=\columnwidth]{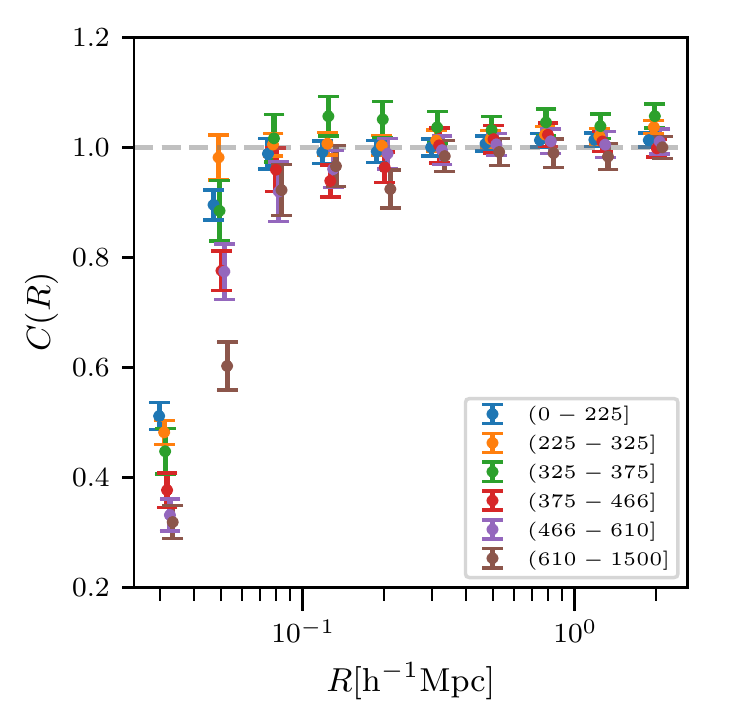}
    \caption{The left (right) panel of the figure shows the boost factors as a function of projected radial position $R(h^{-1} Mpc)$ computed for galaxy groups binned by group luminosity (velocity dispersion). The boost factors are consistent with unity for scales $R>0.1 \hinvMpc$. }
    \label{fig:boost}
\end{figure*}

In order to estimate this dilution, we compute the boost factor using  Eq.~\ref{eqn_7} where we take the ratio of the number of source-lens pairs and source-random pairs. In Figure \ref{fig:boost}, we show our estimate of the boost parameter $C(R)$ as a function of projected radii for bins in both the $r$-band group luminosity and velocity dispersion. The errors come from a large number of random galaxy samples where each sample has the same number of points as our galaxy group catalogues. The grey horizontal line represents a value of unity. 

In the radial range $R > 0.1 \hinvMpc$, the boost factor is consistent with unity, which implies that the contamination from member galaxies is minimal. We believe this to be due to the high median redshift of the HSC source galaxies compared to the redshifts of our lens sample. Therefore we do not include the boost factor correction for the weak lensing signal in this radial range. For smaller distances $R < 0.1 \hinvMpc$, we see evidence of a significant decrease in the boost factor with values of $C(R)$ as small as $0.25$. This implies that we see a deficit in the effective number of lens-source pairs than we expect in the case of random realizations of the lensing sample. Similar behaviour has been reported in the weak lensing study of BOSS galaxies \citep[check figure 7 in][]{2015ApJ...806....1M} and it had been argued that such features are likely due to the presence of bright foreground lenses, which could affect the detection of the source galaxies near galaxy group centres. 

To investigate the impact of such issues on the results in the present analysis, we checked its effect on the mean halo mass derived for all the scaling relations, by restricting our analyses to distances $R > 0.1 \hinvMpc$. We found a negligible impact on the mean halo mass constraints in this reanalysis of our measurements. At small radii, our ESD measurements have large errors, and thus they have a limited impact on our determination of the mean halo masses. Any differences in the measured signal at these scales can also get absorbed by our nuisance parameters and thus have very limited impact on the inference of the mean halo masses.

\section{Comparison with NFW profile based modeling}
\label{app:aum}

The lower limit of $10^{12}\hinvMsun$ on the prior on $\log M'$ adopted in this study (see Table \ref{table_2}) reflects the limits of the resolution to which the \textsc{DARK EMULATOR} was calibrated \citep[][]{2019ApJ...884...29N, 2021arXiv210100113M}. In the case of galaxy groups with the smallest values of group luminosity or velocity dispersion, this modelling threshold could potentially bias the estimates of the mean halo masses (see top left subpanels in Figure. \ref{fig:plogm}). In order to test the impact of such resolution issues, we carried out the modelling using the public python package \textsc{AUM} that uses a NFW profile for the halo matter cross-correlation. We used the same HOD parameterization as that given in Section~\ref{sec_4}, with an uninformative prior on $\log M'$ in the range $[9, 16]$. The package uses the halo mass and halo bias function given by \citet[][]{2010ApJ...724..878T}. Given the uncertainty in the halo mass-concentration relation, it also has an additional free parameter $f_c$, which scales the amplitude of the default halo mass-concentration relation adopted in the code \citep[][]{2007MNRAS.378...55M}.  We use a Gaussian prior on the parameter $f_c$ with a mean of unity and a width of 0.2. This setup is similar to that used in other studies of the galaxy dark matter connection and cosmology \citep[for eg.][]{2015ApJ...806....2M, 2013MNRAS.430..767C}, we show the results of the mean halo masses for both the group $r$-band luminosity and velocity dispersion. We see that the estimates of the mean halo masses $\log \avg{M}$ obtained from \textsc{AUM} are in good agreement with those \textsc{DARK EMULATOR} modelling scheme. This shows that the mass threshold does not cause considerable bias in our estimates of mean halo masses of our galaxy groups.

\begin{figure*}
    \centering
    \includegraphics[width=\columnwidth]{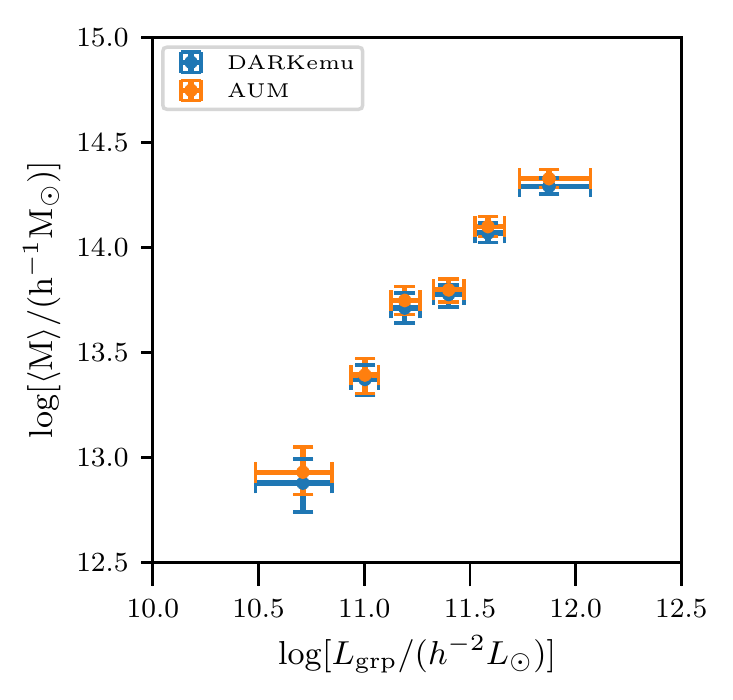}
    \includegraphics[width=\columnwidth]{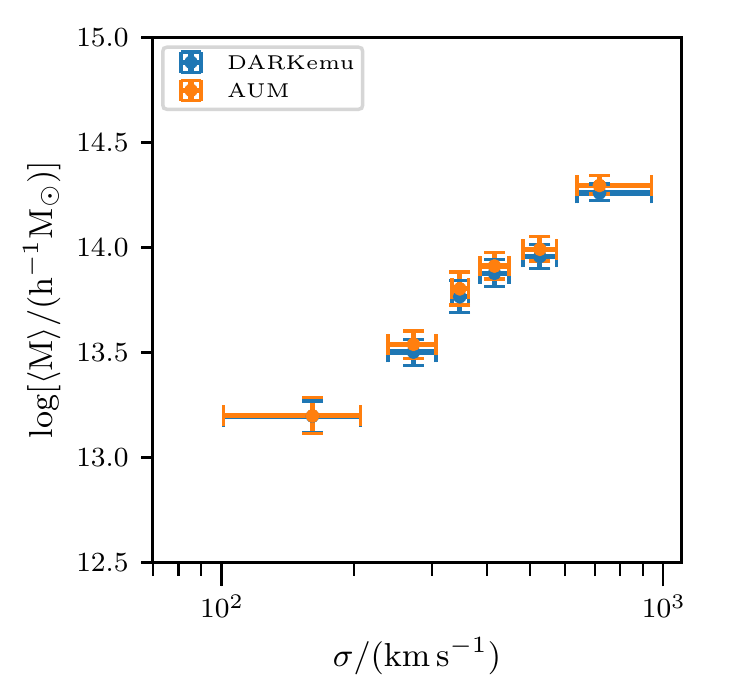}
    \caption{The figure compares the mean halo masses as a function of group luminosity (left panel) and group velocity dispersion (right panel), obtained by using the \textsc{DARK EMULATOR} for carrying out the halo model analysis versus the analytical modeling code \textsc{AUM} which does not have a resolution limitation at the low mass end. We do not see a large impact of the lower limit of $10^{12} \hinvMsun$ on our results. }
    \label{fig:aum_fits}
\end{figure*}

\section{Jackknife errors}
\label{app:jackknife}
In our analysis, we have used shape noise covariances for computing the likelihood of the data given our model parameters. On large scales, the shape noise covariance may underestimate the true covariance. We use the jackknife technique to obtain the covariance. We divided the overlap area between GAMA and HSC into $100$ regions with an unmasked area of about ${\rm 1 deg}^2$ each and recomputed our measurements by dropping each of the regions one at a time. These estimates give us the jackknife covariance. This results in a signal-to-noise of $39$ for group luminosity and $35.8$ for velocity dispersion based group selections.

\begin{figure*}
    \centering
    \includegraphics[width=\columnwidth]{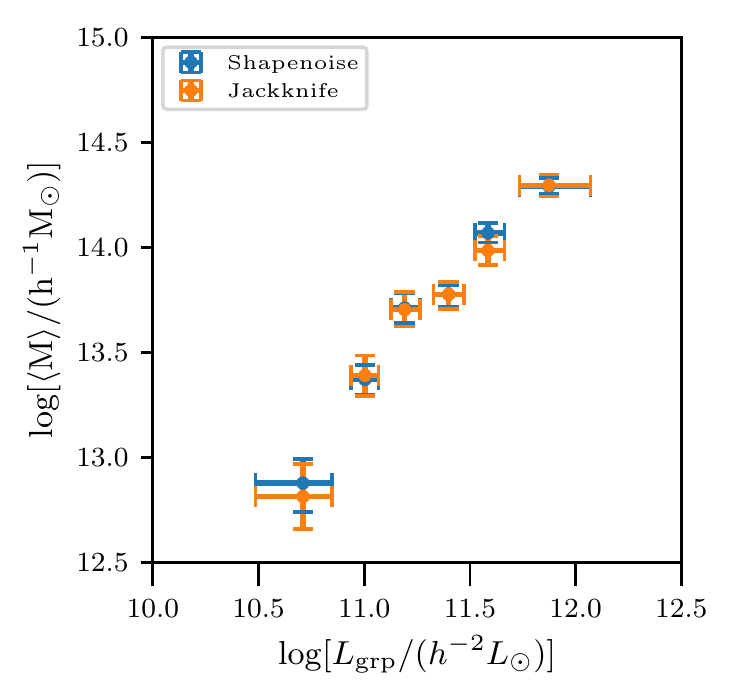}
    \includegraphics[width=\columnwidth]{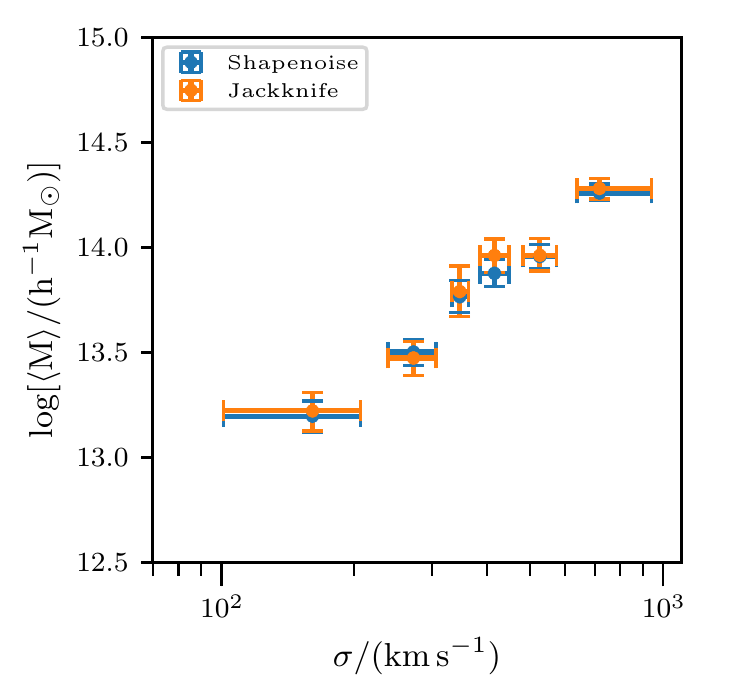}
    \caption{The figure compares the mean halo masses obtained by using the shape noise covariance compared to the covariance obtained from the jackknife technique in the likelihood. We see very little difference in our results between the results obtained from the two different covariance estimation methods.}
    \label{fig:jack_fits}
\end{figure*}

In Figure \ref{fig:jack_fits}, we show the comparison of the mean halo masses inferred with the use of the jackknife covariance and those inferred using the shape noise covariance. The halo masses and the errors agree with each other regardless of our choice the shape noise or jackknife covariances.

We present the comparison between the fits to the scaling relations for both the mass-group luminosity relation and the mass-velocity dispersion relation in Table~\ref{tab:jk_sn_comparison}.

\begin{table*}
\centering
\begin{tabular}{lllllll}
 \hline
Covariance & A & $\alpha$ & $\tilde{A}$ & $\tilde{\alpha}$ & $\sigma_{\rm int}$\\
 \hline
 \hline
\multicolumn{1}{|c|}{} &\multicolumn{6}{|c|}{Mass-Group luminosity}\\ 
\hline
Shape noise
&$0.81\pm0.04$
&$1.01\pm0.07$
&$0.31\pm0.09$
&$0.97\pm0.06$
&$0.58\pm0.10$
&
\\
Jackknife
&$0.78\pm0.05$
&$1.01\pm0.08$
&$0.29\pm0.09$
&$0.98\pm0.07$
&$0.59\pm0.09$
&
\\
\hline
\hline
\multicolumn{1}{|c|}{} &\multicolumn{6}{|c|}{Mass-Group velocity dispersion}\\ 
\hline
Shape noise
&$0.93\pm0.05$
&$1.52\pm0.10$
&$0.35\pm0.10$
&$1.58\pm0.10$
&$0.59\pm0.09$
&
\\
Jackknife
&$0.98\pm0.07$ 
&$1.53\pm0.13$
&$0.35\pm0.10$
&$1.62\pm0.12$
&$0.59\pm0.10$
\\
\hline
\end{tabular}
\caption{Comparison of the posterior distribution of the parameters of the different scaling relations presented in the paper and its dependence on the covariance used in the analysis.}
\label{tab:jk_sn_comparison}
\end{table*}

\bsp	
\label{lastpage}

\end{document}